\documentclass[12pt,preprint]{aastex}

\newcommand{\eg}{{\it e.g.}}
\newcommand{\etal}{et~al.}

\newcommand{\jhk}{$JHK_{\rm s}$}

\newcommand{\ks}{$K_{\rm s}$}

\newcommand{\teff}{$T_{\rm eff}$}
\newcommand{\ubvri}{$UBVR_{\rm C}I_{\rm C}$}

\newcommand{\vsini}{$v \sin i$}
\newcommand{\mum}{$\mu$m}

\begin{document}

\title{Spitzer/MIPS Observations of Stars in the Beta Pictoris
Moving Group}

\slugcomment{Version from \today}

\author{L.\ M.\ Rebull\altaffilmark{1}, 
K.\ R.\ Stapelfeldt\altaffilmark{2}, 
M.\ W.\ Werner\altaffilmark{2}, 
V.\ G.\ Mannings\altaffilmark{1},
C.\ Chen\altaffilmark{3}, 
J.\ R.\ Stauffer\altaffilmark{1},
P.\ S.\ Smith\altaffilmark{5},
I.\ Song\altaffilmark{1},
D.\ Hines\altaffilmark{4},
F.\ J.\ Low\altaffilmark{5}
}

\altaffiltext{1}{Spitzer Science Center/Caltech, M/S 220-6, 1200
E.\ California Blvd., Pasadena, CA  91125
(luisa.rebull@jpl.nasa.gov)}
\altaffiltext{2}{Jet Propulsion Laboratory, California Institute
of Technology, Pasadena, CA 91109}
\altaffiltext{3}{NOAO, P.O.\ Box 26732, Tucson, AZ 85726-6732}
\altaffiltext{4}{Space Science Institute}
\altaffiltext{5}{Steward Observatory, University of Arizona, 933 N.
Cherry Ave., Tucson, AZ 85721}

\begin{abstract}

We present Multiband Imaging Photometer for Spitzer (MIPS)
observations at 24 and 70 \mum\ for 30 stars, and at 160 \mum\ for a
subset of 12 stars, in the nearby ($\sim$30 pc), young ($\sim$12 Myr)
Beta Pictoris Moving Group (BPMG).   In several cases, the new MIPS
measurements resolve source confusion and background contamination
issues in the IRAS data for this sample. We find that 7 members have
24 \mum\ excesses, implying a debris disk fraction of 23\%, and that
at least 11 have 70 \mum\ excesses (disk fraction of $\geq$37\%). 
Five disks are detected at 160 \mum\ (out of a biased sample of 12
stars observed), with a range of 160/70 flux ratios.  The disk
fraction at 24 and 70 \mum, and the size of the excesses measured at
each wavelength, are both consistent with an ``inside-out'' infrared
excess decrease with time, wherein the shorter-wavelength excesses
disappear before longer-wavelength excesses, and consistent with the
overall decrease of infrared excess frequency with stellar age, as
seen in Spitzer studies of other young stellar groups.  Assuming that
the infrared excesses are entirely due to circumstellar disks, we
characterize the disk properties using simple models and fractional
infrared luminosities.  Optically thick disks, seen in the younger TW
Hya and $\eta$ Cha associations, are entirely absent in the BPMG. 

Additional flux density measurements at 24 and 70 \mum\ are reported
for nine Tucanae-Horologium Association member stars.  Since this is
$<$20\% of the association membership, limited analysis on the
complete disk fraction of this association is possible. 

\end{abstract}


\section{Introduction}
\label{sec:intro}

In recent years, several nearby ($\lesssim$100 pc) young
($\lesssim$200 Myr) stellar associations have been identified.  These
groupings provide a special opportunity to study ``up close'' the
evolution of circumstellar material at a potentially crucial phase of
disk evolution, namely that epoch when planets are thought to be
forming.  With the advent of the {\em Spitzer} Space Telescope (Werner
\etal\ 2004), specifically the Multiband Imaging Photometer for
Spitzer (MIPS; Rieke \etal\ 2004), astronomers now can easily study
the properties of the stars in those nearby groupings at much lower
disk excess levels than was possible with 2MASS or ISO.  In some
cases, these stars are close enough, and the disks big enough, that
one can spatially resolve the disk structure, providing even more
information about the disk properties.  While it is thought that all
stars start with massive, optically thick, primordial disks, older
stars possess much less massive, optically thin, second-generation,
debris disks, where the dust to primary star luminosity ratio $L_{\rm
dust}/L_{*} \lesssim 10^{-3}$ (see, e.g., Meyer \etal\ 2007 and
references therein). In this phase, it is thought that
planetesimal-mass bodies have already formed in the disk; collisions
of these bodies can replenish the dust in those systems. The evolution
of mid- to far-infrared-emitting dust grains a few microns in size
within debris disks has been a subject of much study (see, e.g.,
Bryden \etal\ 2006, Su \etal\ 2006, Chen \etal\ 2005, Werner \etal\
2006 and references therein).  The measurement of the overall disk
fraction in clusters of known age (and ultimately measurement of the
dust distribution in individual systems via direct imaging) is key to
understanding disk evolution and planet formation.

The disk around $\beta$ Pictoris has been known since the mid 1980s
when it was one of the first debris disks discovered by the Infrared
Astonomy Satellite (IRAS) mission (Gillett 1986, Paresce \& Burrows
1987).  Little was known about $\beta$ Pic when its infrared excess
was discovered.  It was not located within an obvious star-forming
region or cluster, and even its age was poorly constrained.  New
observations in recent years have placed  $\beta$ Pic in better
context.  A number of other stars have been discovered that share 
$\beta$ Pic's space motion and are believed to be coeval with  $\beta$
Pic (\eg, Barrado y Navascues \etal\ 1999; Zuckerman \etal\ 2001 and
references therein).  At only $\sim$30 pc away with an age of $\sim$12
Myr, this so-called Beta Pic Moving Group (BPMG) is the nearest
identified young stellar association, and has been studied
intensively.  Zuckerman \& Song (2004) and subsequent authors have
identified 30 BPMG member or potential member stellar systems.  

This study presents MIPS observations at 24, 70, and 160 \mum\ of all
of the currently-known BPMG members, as well as several members from
the Tucanae-Horologium Association, another nearby ($\sim$50 pc) young
($\sim$30 Myr) group (Zuckerman \& Song 2004).   We first present the
observational details (\S\ref{sec:obs}), and then discuss
identification of stars with infrared excesses
(\S\ref{sec:findingdisks}). We fit some simple models in
\S\ref{sec:models} to characterize the disk properties for the stars
we have found with excesses. Finally, we discuss the sample as a whole
in \S\ref{sec:disc}, and summarize our conclusions in
\S\ref{sec:concl}.

\clearpage
\begin{deluxetable}{lllllllllll}
\tablecaption{Nearby young association members in this study}
\label{tab:targets}
\tablewidth{0pt}
\tabletypesize{\scriptsize}
\rotate
\tablehead{
\colhead{Ass'n\tablenotemark{a}} & \colhead{HIP} & \colhead{HD} &
\colhead{HR} & \colhead{GJ} & \colhead{other name}
& \colhead{name used here} & \colhead{distance} &
\colhead{spectral} &\colhead{$V$}&\colhead{\ks} \\
& \colhead{number} & \colhead{number} & \colhead{number} &
\colhead{number} & & & \colhead{(pc)} & \colhead{type} &
\colhead{(mag)} & \colhead{(mag)}} 
\startdata
BPMG	&	560	&	203	&	9	&		&		&	HR 9	&	39.1	&	F2 IV	& 6.2 &	5.24	\\
BPMG	&	10679	&	\tablenotemark{b}	&		&		&		&	HIP 10679	&	34.0	&	G2 V	&7.8&	6.26	\\
BPMG	&	10680	&	14082	&		&		&		&	HD 14082	&	39.4	&	F5 V	&7.0&	5.79	\\
BPMG	&	11437	&		&		&		&	AG Tri A	&	AG Tri A	&	42.3	&	K8	&10.1&	7.08	\\
BPMG	&	11437	&		&		&		&	AG Tri B	&	AG Tri B	&	42.3	&	M0	&\ldots&	7.92	\\
BPMG	&	12545	&		&		&		&	BD 05d378	&	HIP 12545	&	40.5	&	M0	&10.4&	7.07	\\
BPMG	&	21547	&	29391	&	1474	&		&	51 Eri	&	51 Eri	&	29.8	&	F0 V	&5.2&	4.54	\\
BPMG	&		&		&		&	3305	&		&	GJ 3305	&	29.8	&	M0.5	&10.6&	6.41	\\
BPMG	&	23309	&		&		&		&	CD-57d1054	&	HIP 23309	&	26.3	&	K8	&10.1&	6.24	\\
BPMG	&	23418	&		&		&	3322	&		&	GJ 3322 A/B	&	32.1	&	M3 V	&11.7&	6.37	\\
BPMG	&	25486	&	35850	&	1817	&		&		&	HR 1817	&	26.8	&	F7/8 V	&6.3&	4.93	\\
BPMG	&	27321	&	39060	&		&		&	$\beta$ Pic	&	Beta Pic	&	19.3	&	A5 V	&3.9&	3.53	\\
BPMG	&	29964	&	45081	&		&		&	AO Men	&	AO Men	&	38.5	&	K7 	&9.9&	6.81	\\
BPMG	&	76629	&	139084	&		&		&	V343 Nor A	&	V343 Nor A/B	&	39.8	&	K0 V	&8.2&	5.85	\\
BPMG	&	79881	&	146624	&	6070	&		&		&	HR 6070	&	43.1	&	A0 (V)	&4.8&	4.74	\\
BPMG	&	84586	&	155555	&		&		&	V824 Ara A/B    &	V824 Ara A/B    &	31.4	&	K1 VP	&6.9&	4.70	\\
BPMG	&	84586	&	155555	&		&		&	V824 Ara C	&	V824 Ara C	&	31.4	&	M4.5	&12.7& 7.63	\\
BPMG	&	88399	&	164249	&		&		&		&	HD 164249	&	46.9	&	F5 V	&7.0&	5.91	\\
BPMG	&	88726A	&	165189	&	6749	&		&		&	HR 6749/HR 6750	&	43.9	&	A5 V	&5.0&	4.39	\\
BPMG	&	92024	&	172555	&	7012	&		&		&	HR 7012	&	29.2	&	A5 IV/V	&4.8&	4.30	\\
BPMG	&		&		&		&		&	CD-64D1208AB	&	CD-64D1208 A/B	&	29.2	&	K7 	&10.4&	6.10	\\
BPMG	&	92680	&	174429	&		&		&	PZ Tel	&	PZ Tel	&	49.7	&	K0 VP	&8.4&	6.37	\\
BPMG	&	95261	&	181296	&	7329	&		&	$\eta$ Tel 	&	eta Tel A/B	&	47.7	&	A0 V	&5.1&	5.01	\\
BPMG	&	95270	&	181327	&		&		&		&	HD 181327	&	50.6	&	F5/6 V	&7.0&	5.91	\\
BPMG	&	102141	&	196982	&		&	799	&	AT Mic 	&	AT Mic A/B	&	10.2	&	M4.5	&10.3&	4.94	\\
BPMG	&	102409	&	197481	&		&	803	&	AU Mic	&	AU Mic	&	9.9	&	M1 Ve	&8.8&	4.53	\\
BPMG	&	103311	&	199143	&		&		&		&	HD199143 A/B	&	47.7	&	F8 V	&7.3&	5.81	\\
BPMG	&		&	358623	&		&		&	AZ Cap A, BD-17d6128	&	AZ Cap A/B	&	47.7	&	K7/M0	&10.6&	7.04	\\
BPMG	&	112312	&		&		&		&	WW PsA A	&	WW PsA A	&	23.6	&	M4e	&12.2&	6.93	\\
BPMG	&	112312	&		&		&		&	WW PsA B	&	WW PsA B	&	23.6	&	M4.5e	&13.4&	7.79	\\
\hline
Tuc-Hor	&	1113	&	987	&		&		&		&	HD 987	&	43.7	&	G6 V	& 8.7&	6.96	\\
Tuc-Hor	&	3556	&		&		&		&		&	HIP 3556	&	38.5	&	M3	& 12.3&	7.62	\\
Tuc-Hor	&		&		&		&		&	CPD-64d120	&	CPD-64d120	&	29.2	&	K7 	& 9.5&	8.01	\\
Tuc-Hor	&	7805	&	10472	&		&		&		&	HD 10472	&	66.6	&	F2 IV/V	&7.6 & 	6.63	\\
Tuc-Hor	&	9685	&	12894	&		&		&		&	HD 12894	&	47.2	&	F2 V 	&6.4 & 	5.45	\\
Tuc-Hor	&	10602	&	14228	&	674	&		&	$\phi$ Eri	&	Phi Eri	&	47.5	&	B8 V	&	3.6& 4.13	\\
Tuc-Hor	&		&		&		&		&	GSC 8056-0482	&	GSC 8056-0482	&	30.9	&	M3 Ve	&12.1&	7.50	\\
Tuc-Hor	&	12394	&	16978	&	806	&		&	$\epsilon$ Hya	&	Eps Hya	&	47.0	&	B9 V	&4.1&	4.25	\\
Tuc-Hor	&	101612	&	195627	&	7848	&		&		&	HR 7848	&	27.6	&	F0 V 	&4.8&	4.04	\\
\enddata
\tablenotetext{a}{The abbreviations ``BPMG" and ``Tuc-Hor" denote
the Beta Pic Moving Group and the Tucanae-Horologium associations, respectively.}
\tablenotetext{b}{Some references list this object as HD 14082B.}
\end{deluxetable}
\clearpage
\thispagestyle{empty}
\begin{deluxetable}{cllllrrr}
\tablecaption{Summary of observations}
\label{tab:observations}
\tablewidth{0pt}
\rotate
\tabletypesize{\scriptsize}
\tablehead{
\colhead{Spitzer} & \colhead{PI} & \colhead{AORKEY} & \colhead{date of
observation} & 
\colhead{object(s)} &  \colhead{24 \mum\ integ.} & \colhead{70 \mum\ integ.} & 
\colhead{160 \mum\ integ.} \\
\colhead{program id} & & \colhead{(Spitzer Archive identifier)} && & \colhead{time (s)} & \colhead{time (s)}  & \colhead{time (s)}}
\startdata
102	& Werner  &  9018624, 9020928 & 2005-12-01, 2004-06-21&       HR 9    &       48      &       126     &       63      \\
3600	& Song    &  11256064 &2005-01-29&       HD 14082, HIP 10679     &       48      &       231     &       0       \\
102	& Werner  &  9019392, 9020672&2005-08-29, 2006-02-18&       AG Tri A/B      &       48      &       231     &       63      \\
102	& Werner  &  9017600, 9020160&2005-09-05, 2006-02-17&  HIP 12545       &       48      &       231     &       63      \\
102	& Werner  &  9019904 &2004-09-16& 51 Eri, GJ 3305 &       48      &       231     &       0       \\
102	& Werner  &  9024000 &2004-11-08& HIP 23309       &       48      &       126     &       0       \\
102	& Werner  &  9023744 &2004-10-14& GJ 3322 A/B     &       48      &       126     &       0       \\
148	& FEPS    &  5252352, 5446656, 5447424& 2005-02-27 (all 3) &     HR 1817 &       276     &       693     &       84      \\
80	& Werner  &  8970240, 12613632, 4884992& 2004-03-20, 2005-04-04, 2004-02-21& $\beta$ Pic        &       36      &       252     &       27      \\
148	& FEPS    &  5222656 & 2004-10-13&      AO Men  &       92      &       881     &       84      \\
148	& FEPS    &  5223424 & 2004-07-29 &       V343 Nor A/B    &       92      &       231     &       0       \\
10	& Jura    &  3720704 & 2004-02-25 &     HR 6070 &       48      &       126     &       0       \\
84	& Jura    &  4813056 & 2004-09-17&       V824 Ara A/B/C  &       48      &       126     &       0       \\
102	& Werner  &  9019136, 9023488& 2005-08-30, 2004-03-17 & HD 164249       &       48      &       101     &       63      \\
102	& Werner  &  9023232 & 2004-09-23&      HR 6749/HR 6750 &       48      &       231     &       0       \\
10	& Jura    &  3723776 & 2004-04-07&      HR 7012, CD-64D1208A/B  &       48      &       126     &       0       \\
72	& Low     &  4554496 & 2004-04-06&      PZ Tel  &       180     &       545     &       0       \\
57	& Rieke   &  8934912, 8935168, 8935424 &2004-04-09, 2004-04-09, 2004-04-07    &       $\eta$ Tel &       48      & 100   &       76      \\
72	& Low     &  4556032 & 2004-04-06 &     HD 181327       &       92      &       126     &       63      \\
80	& Werner  &  4637184 & 2004-05-11&      AT Mic A/B      &       48      &       231     &       0       \cr
3657	& Graham  &  11403008& 2005-05-20&      AU Mic (160 only)       &       0       &       0       &       629     \cr
80	& Werner  &  4637440 & 2004-05-02&     AU Mic  &       48      &       231     &       0       \cr
148	& FEPS    &  5254656 & 2004-10-13&      HD 199143 A/B   &       92      &       231     &       84      \cr
80	& Werner  &  4643840 & 2004-05-11&      AZ Cap A/B      &       48      &       545     &       0       \cr
102	& Werner  &  9021696 & 2004-05-31&     WW PsA A/B      &       48      &       126     &       0       \\
\hline
102	& Werner  &  9022976 & 2004-11-05&HD 987  &       48      &       126     &       0       \\
102	& Werner  &  9022720 &2004-11-03&       HIP 3556        &       48      &       231     &       0       \\
102	& Werner  &  9022464 &2004-05-11&       CPD-64d120      &       48      &       0       &       0       \\
102	& Werner  &  4945152 &2004-11-07&       HD 10472        &       48      &       69      &       0       \\
102	& Werner  &  9022208 &2004-11-07&       HD 12894        &       48      &       126     &       0       \\
102	& Werner  &  9021952 &2004-11-08&       $\phi$ Eri &       48      &       231     &       0       \\
102	& Werner  &  9021440 & 2004-11-07&     GSC 8056-482    &       48      &       126     &       0       \\
102	& Werner  &  9021184 & 2005-06-19&  $\epsilon$ Hya &       48      &       231     &       0       \\
102	& Werner  &  9020416 & 2004-04-13&     HR 7848 &       48      &       69      &       0       \\
\enddata
\end{deluxetable}
\clearpage

\begin{deluxetable}{lcccccc}
\tablecaption{Results: MIPS flux densities}
\label{tab:results}
\tablewidth{0pt}
\tabletypesize{\scriptsize}
\tablehead{
\colhead{object} & \colhead{photospheric} &\colhead{measured} 
& \colhead{photospheric} &\colhead{measured}
& \colhead{photospheric} &\colhead{measured}\\
& \colhead{24 \mum\ (mJy)} &\colhead{24 \mum\ (mJy)\tablenotemark{a}} 
& \colhead{70 \mum\ (mJy)} 
& \colhead{70 \mum\ (mJy)\tablenotemark{b}} 
& \colhead{160 \mum\ (mJy)} &\colhead{160 \mum\
(mJy)\tablenotemark{c}} }
\startdata
           HR 9&   60&  109\tablenotemark{d}&  7.0&$ $ 61\tablenotemark{d}& 1.3&$<$ 27 \\
      HIP 10679&   23&   39\tablenotemark{d}&  2.7& 43.0\tablenotemark{d}& 0.5&$ $\ldots\\
      HD 14082 &   36&   37                 &  4.2&$<$ 18                & 0.8&$ $\ldots\\
       AG Tri A&   14&   17                 &  1.7&$ $ 75.1\tablenotemark{d}& 0.3&$<$ 35 \\
       AG Tri B&  7.2&  7.1                 &  0.9&$<$  23                & 0.2&$<$ 35 \\
      HIP 12545&   16&   12                 &  1.9&$<$ 25                & 0.4&$<$ 50 \\
         51 Eri&  114&  115                 & 13.2&$<$ 23                & 2.5&$ $\ldots\\
        GJ 3305&   29&   24                 &  3.4&$<$ 23                & 0.6&$ $\ldots\\
      HIP 23309&   34&   27                 &  4.0&$<$ 24                & 0.8&$ $\ldots\\
    GJ 3322 A/B&   30&   28                 &  3.8&$<$ 39                & 0.7&$ $\ldots\\
        HR 1817&   79&   79                 &  9.2&$ $ 44.7\tablenotemark{d}& 1.7&$<$ 77 \\
$\beta$ Pic\tablenotemark{e}& 280& 7276\tablenotemark{d}&  32&$ $12990\tablenotemark{d}& 5.9& 3646\tablenotemark{d}\\
         AO Men&   15&   15                 &  1.7&$<$  8                & 0.3&$<$ 28 \\
   V343 Nor A/B&   34&   34                 &  4.0&$<$ 86                & 0.7&$ $\ldots\\
        HR 6070&   90&   97                 & 10.4&$<$ 77                & 1.9&$ $\ldots\\
   V824 Ara A/B&  100&   97                 & 11.9&$<$ 25                & 2.2&$ $\ldots\\
     V824 Ara C&  9.9&   11                 &  1.3&$<$ 25                & 0.2&$ $\ldots\\
      HD 164249&   32&   76\tablenotemark{d}&  3.7&$ $624\tablenotemark{d}& 0.7& 104\tablenotemark{d} \\
HR 6749/HR 6750&  120&  113                 & 14.4&$<$ 27                & 2.7&$ $\ldots\\
        HR 7012&  130&  766\tablenotemark{d}& 15.6&$ $197\tablenotemark{d}& 2.9&$ $\ldots\\
 CD-64D1208 A/B&   35&   30                 &  4.2&$<$ 23                & 0.8&$ $\ldots\\
         PZ Tel&   21&   21                 &  2.5&$ $ 17.4\tablenotemark{d}& 0.5&$ $\ldots\\
$\eta$ Tel A/B\tablenotemark{e}&   70&  382\tablenotemark{d}&  8.1&$ $409\tablenotemark{d}& 1.5&$ $ 68\tablenotemark{d}\\
      HD 181327&   32&  195\tablenotemark{d}&  3.7&$ $1468\tablenotemark{d}& 0.7&$ $ 658\tablenotemark{d} \\
     AT Mic A/B&  118&  116                 & 14.9&$<$ 18                & 2.9&$ $\ldots\\
         AU Mic&  164&  143                 & 19.4&$ $205\tablenotemark{d}& 3.7&$ $ 168\tablenotemark{d} \\
  HD 199143 A/B&   35&   35                 &  4.0&$<$ 22                & 0.8&$<$  31 \\
     AZ Cap A/B&   15&   13                 &  1.8&$<$ 12                & 0.3&$ $\ldots\\
       WW PsA A&   19&   18                 &  2.4&$<$ 27                & 0.5&$ $\ldots\\
       WW PsA B&  8.6&  9.1                 &  1.1&$<$ 27                & 0.2&$ $\ldots\\
\hline
         HD 987&   12&   12                 &  1.4&$<$  21                & 0.3&$ $\ldots\\
       HIP 3556&  9.5&  8.4                 &  1.2&$<$  16                & 0.2&$ $\ldots\\
     CPD-64d120&  6.1&  4.9                 &  0.7&$ $\ldots              & 0.1&$ $\ldots\\
       HD 10472&   17&   26\tablenotemark{d}&  1.9&$ $127\tablenotemark{d}& 0.4&$ $\ldots\\
       HD 12894&   49&   46                 &  5.8&$<$ 20                & 1.1&$ $\ldots\\
     $\phi$ Eri&  150&  170                 & 17.3&$<$ 17                & 3.2&$ $\ldots\\
  GSC 8056-0482&   11&  9.0                 &  1.3&$<$  24                & 0.3&$ $\ldots\\
     $\epsilon$ Hya&  140&  124                 & 15.9&$ $ 12.6                & 3.0&$ $\ldots\\
        HR 7848&  179&  186                 & 20.8&$ $609\tablenotemark{d}& 3.9&$ $\ldots\\
\enddata
\tablenotetext{a}{The systematic uncertainty on our 24 \mum\ flux
densities is estimated to be 4\% (Engelbracht \etal\ 2008).}
\tablenotetext{b}{The systematic uncertainty on our 70 \mum\ flux
densities is estimated to be 10\% (Gordon \etal\ 2008; see also
discussion in text).  Upper limits quoted here are 3-$\sigma$.}
\tablenotetext{c}{The systematic uncertainty on our 160 \mum\ flux
densities is estimated to be 12\% (Stansberry \etal\ 2008).  Upper
limits quoted here are 3-$\sigma$.}
\tablenotetext{d}{Infrared excess and inferred disk at this
wavelength; see text for discussion as to how these disk candidates
were selected.}
\tablenotetext{e}{The 24 and 70 \mum\ flux densities are from Su
\etal\ (2006); see text.}
\end{deluxetable}
\clearpage

\begin{figure*}[tbp]
\epsscale{0.8}
\plotone{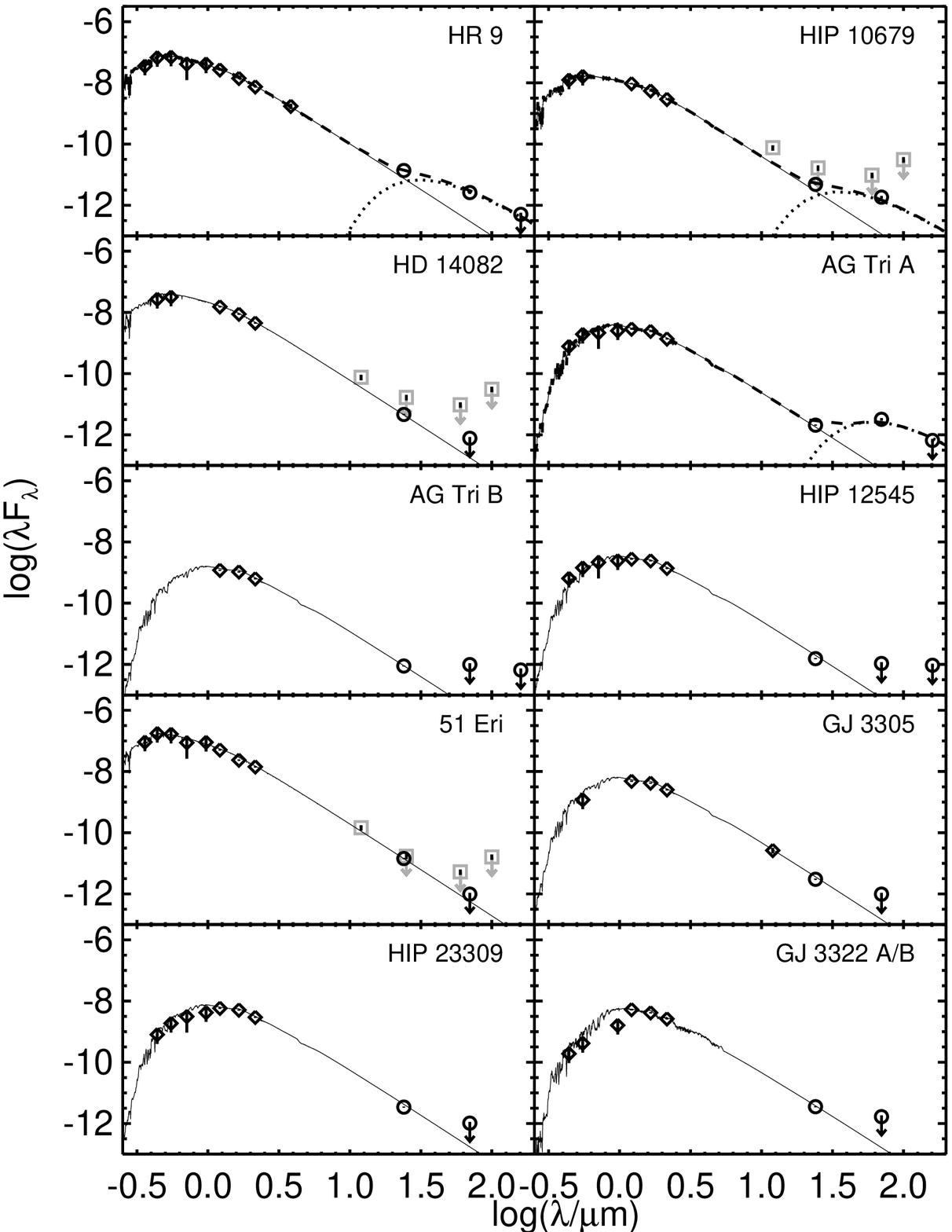}
\caption{Spectral Energy Distributions (SEDs) for all the BPMG targets
discussed in this paper, part 1.  The $x$-axis plots log of the wavelength in
microns, and the $y$ axis plots log($\lambda F_{\lambda}$) in cgs
units (ergs s$^{-1}$ cm$^{-2}$).  Points gleaned from the literature
are diamonds, boxes are detections or upper limits from IRAS, and
circles are new MIPS points.  Downward-pointing arrows indicate upper
limits.  The stellar model that is plotted is selected from the
Kurucz-Lejeune grid (see text for discussion), normalized
to the \ks\ band as observed; the simple disk model shown here is
described in the text.  The dotted line is the disk component alone,
and the dashed line is the sum of the disk plus star model when more
than one data point describes the disk.}
\label{fig:seds1}
\end{figure*}

\begin{figure*}[tbp]
\epsscale{0.8}
\plotone{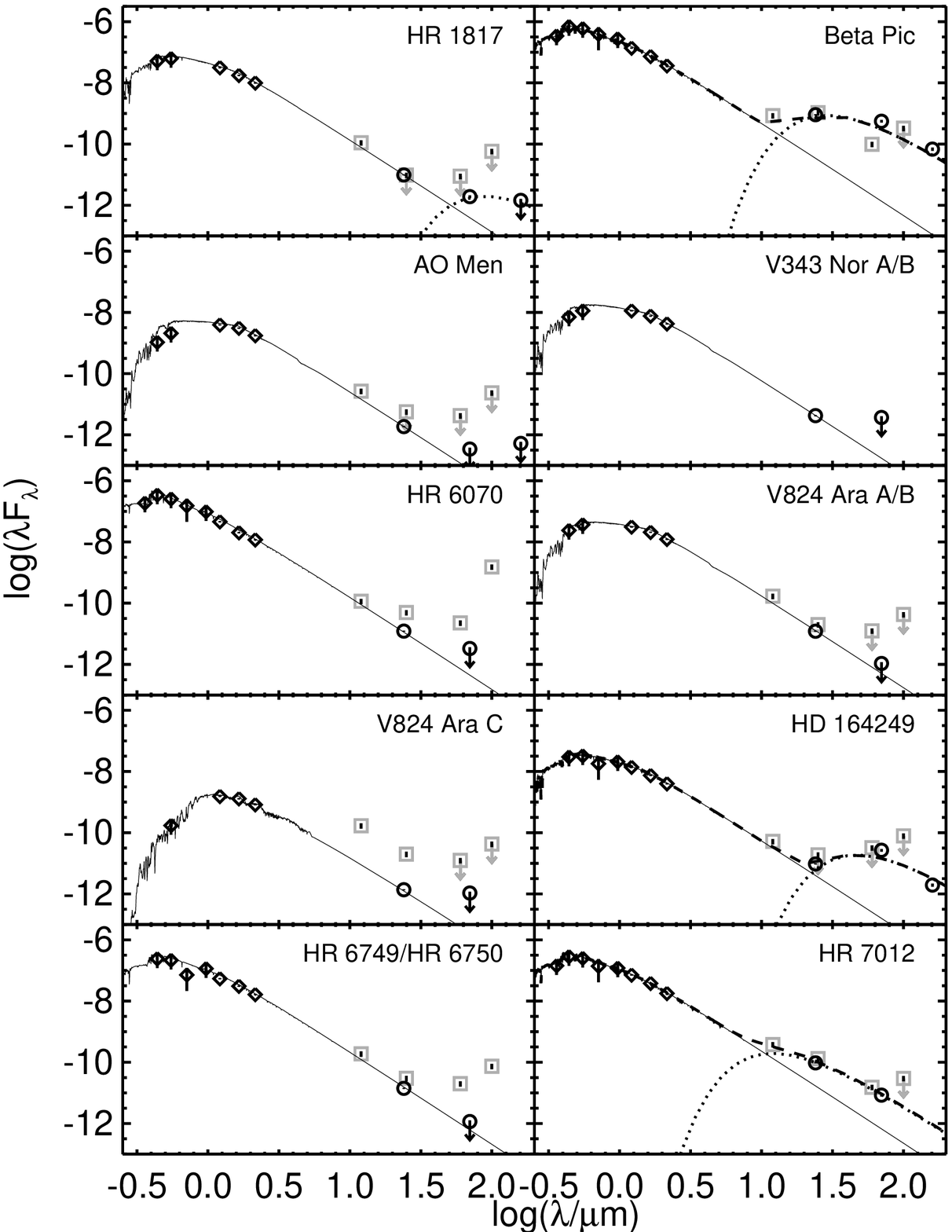}
\caption{SEDs for all the BPMG targets discussed in this paper, part 2.}
\label{fig:seds2}
\end{figure*}

\begin{figure*}[tbp]
\epsscale{0.8}
\plotone{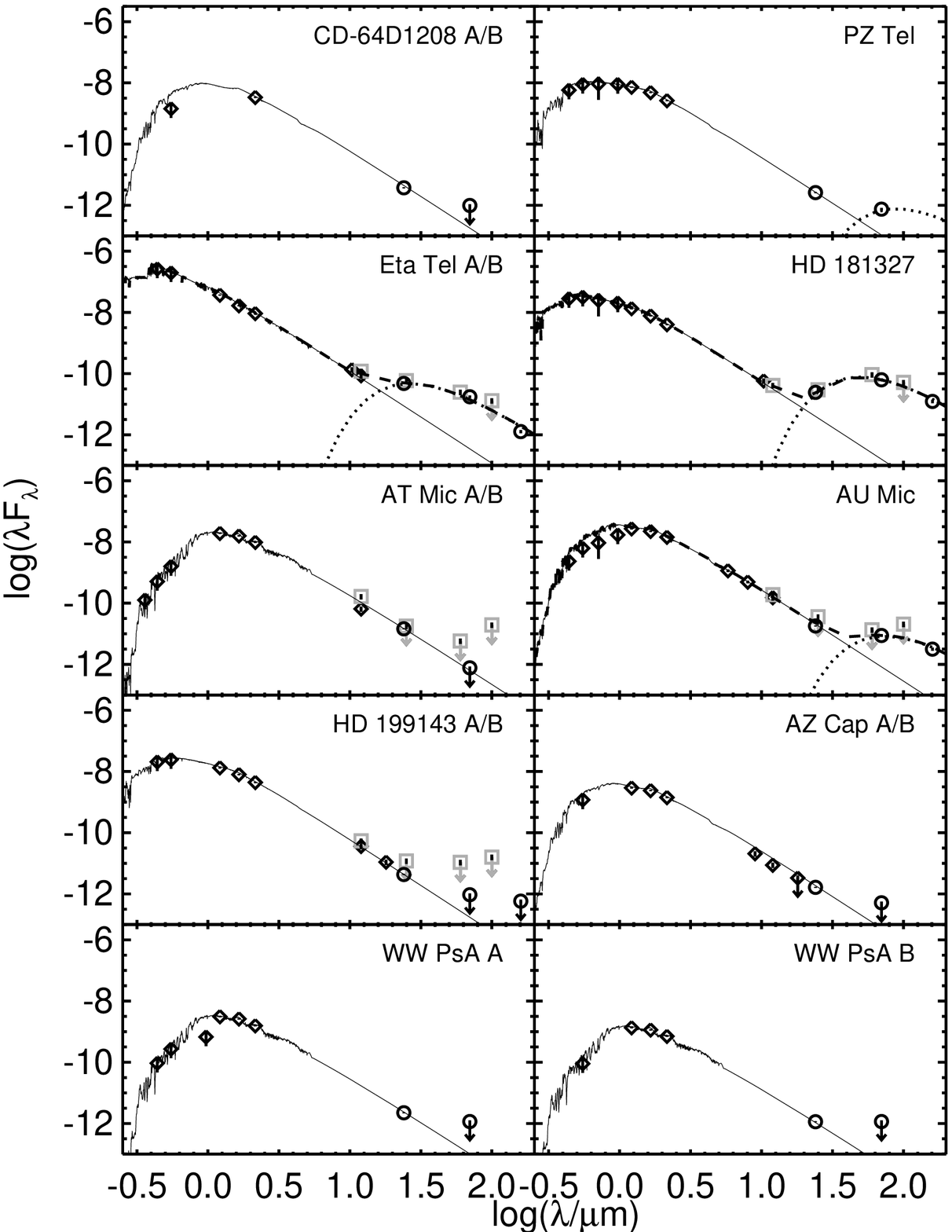}
\caption{SEDs for all the BPMG targets discussed in this paper, part 3.}
\label{fig:seds3}
\end{figure*}

\begin{figure*}[tbp]
\epsscale{0.8}
\plotone{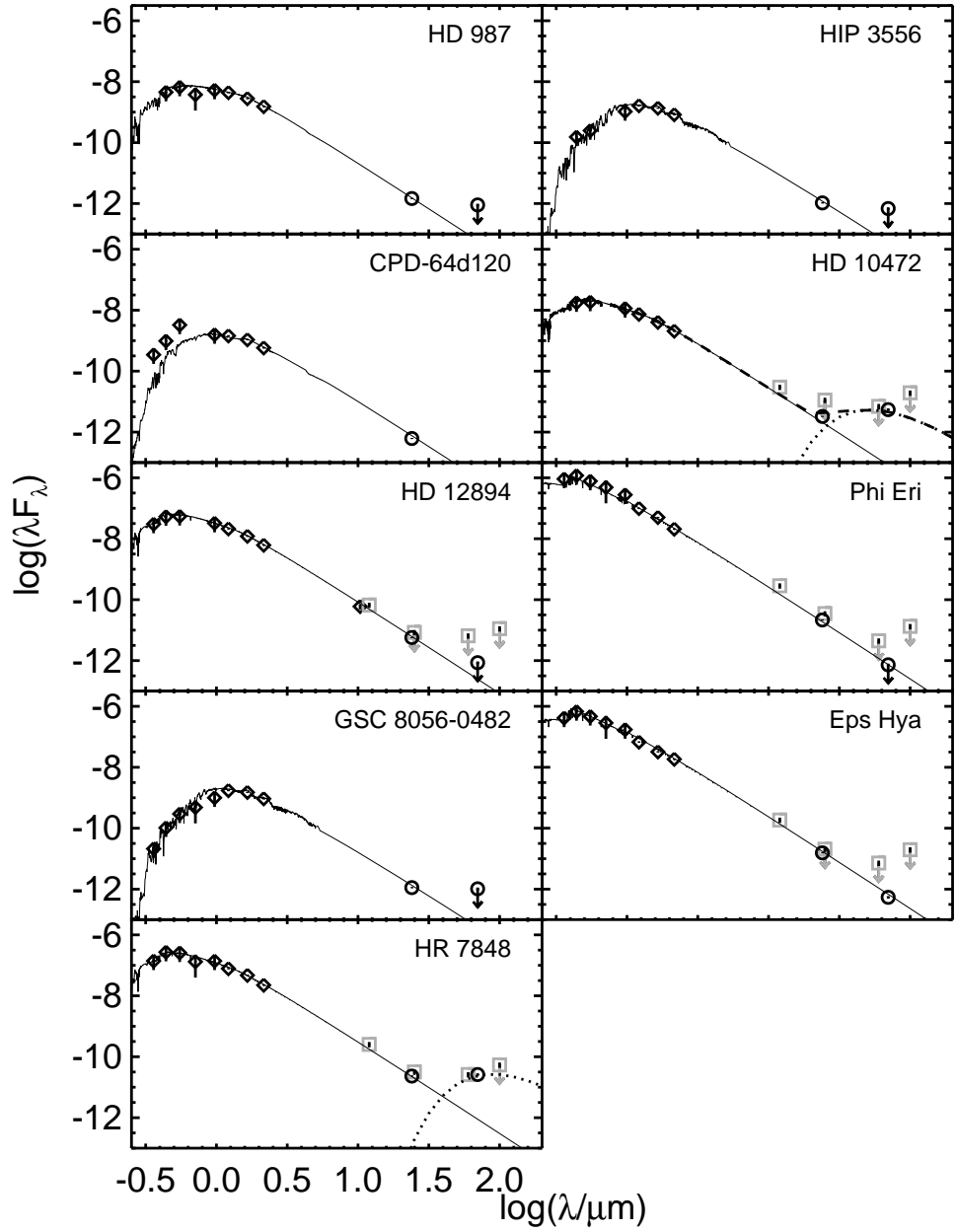}
\caption{SEDs for all the Tuc-Hor targets discussed in this paper. 
Notation is as in previous plots.}
\label{fig:seds4}
\end{figure*}
\clearpage

\section{Observations, Data Reduction, and Ancillary Data}
\label{sec:obs}
\subsection{Target selection and observations}

Many individual member stars from nearby young stellar clusters are
scattered among several of the programs originating with the Spitzer
guaranteed-time observers (GTO).  The Spitzer GTO program 102
(P.I.-M.\ Werner) observed 13 BPMG stars or systems, along with nine
stars (or systems) thought to be Tucanae-Horologium association
members (where membership for both associations is as reported by
Zuckerman \& Song 2004).  As the observers of record for this program,
we felt it important to report the observations for all of the stars
from this program.  In order to enhance the discussion, we assembled a
list of all of the stars or stellar systems thought to be members of
the BPMG, based on Zuckerman \& Song (2004).  We retrieved data for
the remaining stars/star systems of the BPMG out of the Spitzer
Archive.  Nearly all of these observations come from GTO programs and
were obtained over the first 3 years of the mission.

The 39 stars or systems discussed in this paper -- all the targets
from program 102, plus the remaining Beta Pic member stars from the
archive -- are listed in Table~1, along with cluster membership.  Note
that binary systems unresolved by MIPS are listed together, e.g, GJ
3322 A/B, and that these unresolved binaries are effectively treated
as single stars throughout the rest of this paper. The implications of
this decision will be discussed further below. The Tuc-Hor stars are
included at the bottom of this table (and the next two tables),
separated from the BPMG stars by a line.

All of the Spitzer archive identifications (AORKEYs) and other
assorted program information (including program IDs and dates of
observation) are listed in Table~2.  Since the observations were
acquired from a variety of programs, the integration times used for
each target are not uniform (see Table~2).  All targets were observed
at 24 \mum, all but one at 70 \mum\ (CPD-64d120), and a subset of
twelve sources, not selected uniformly, were observed at 160 \mum. All
objects are detected at good signal-to-noise ($>$25) at 24 \mum; there
are many upper limits at both 70 and 160 \mum. 

Since we assembled our list of members from Zuckerman \& Song (2004),
we are obviously missing any undiscovered members. Thus, we cannot
assert that our study is complete over all possible BPMG members.  As
discussed in Song \etal\ (2008) and Torres \etal\ (2006), surveys of
young stars near Earth form distinctive groupings in
age-velocity-position space, and young members earlier than M are
easily identified via, e.g., lithium absorption.  It is unlikely that
there are many undiscovered members earlier than M. However, it is
possible that there are undiscovered M star members.  These
intrinsically fainter M stars often lack parallax measurements, and
moreover are often harder to confirm as members, because they, for
example, deplete Li faster than higher-mass stars. 

Because the BPMG is physically close to us, it subtends a large angle
on the sky, and finding additional members often requires searching
over a large area.  However, co-moving companions may be found in
close proximity to known members, as evidenced by the number of known
companions in the BPMG.  Our observations cover relatively small
regions around each member star, so we have a chance of finding  these
sorts of close companions.  Several observations detect additional
objects in the field near BPMG members; since these objects are bright
enough to be detected in these shallow observations, these objects
could also be potential association members, and the argument for BPMG
membership might be made if these objects have infrared excesses.  We
examined our data for any additional stars with excesses
serendipitously included within the Spitzer field of view, but none
were detected; see the Appendix for discussion of each individual
case.

\subsection{Data reduction}

All of the observations were conducted in MIPS photometry mode.  Most
of the observations were conducted using the small field photometry
astronomical observing template (AOT) and, at 70 \mum, the default
pixel scale.  For these targets (including those in the literature),
we reprocessed the data in a uniform manner in order to limit
systematics introduced by slightly different reduction methods.   Two
objects, $\eta$ Tel and $\beta$ Pic, were observed using  observing
strategies designed for extended objects (e.g., customized sub-pixel
dithering, and 70 \mum\ fine scale).  We note for completeness that,
while $\beta$ Pic is well-resolved at MIPS wavelengths, $\eta$ Tel is
not.  Because these data must be handled differently anyway, rather
than reprocessing the data, we use the MIPS photometry at 24 and 70
\mum\ as reported by Su \etal\ (2006) for both $\eta$ Tel and $\beta$
Pic.  

We started with the Spitzer Science Center (SSC) pipeline-produced
basic calibrated data (BCDs), version S14.  (For a description of the
pipeline, see Gordon \etal\ 2005.)  Since we treated each MIPS channel
differently, each is discussed separately below.

Our detections and upper limits are listed in Table~3.  Note that,
while every target was detected at 24 \mum, one target was not
observed at 70 \mum\ (CPD-64d120), and more than half the sample was
not observed at 160 \mum.   For most of our sample, this is the first
time that MIPS fluxes have appeared in the literature. For the four
FEPS stars that are part of the FEPS final delivery catalog (available
on the SSC website) and for the three stars reported in Chen \etal\
(2005), our fluxes are consistent within the reported errors.

\subsubsection{24 \mum}

All targets were observed at 24 \mum.  For each observation, we
constructed a 24 \mum\ mosaic from the pipeline BCDs using the SSC
mosaicking and point-source extraction (MOPEX) software (Makovoz \&
Marleau 2005), with a pixel scale of 2.5$\arcsec$ px$^{-1}$, close to
the native pixel scale of $2.49\arcsec\times2.60\arcsec$.  We
extracted sources from the 24 \mum\ mosaics using the astronomical
point-source extraction (APEX) 1-frame portion of MOPEX, with point
response function (PRF)-fitting photometry of the image mosaics.  All
of our targets were detected at good signal to noise ($>$25) at 24
\mum.  The systematic uncertainty in the zero-point of the conversion
from instrumental units to calibrated flux units is estimated to be
4\% (Engelbracht \etal\ 2008); the statistical error is much smaller,
and so is not tabulated.

\subsubsection{70 \mum}

At 70 \mum, the SSC pipeline produces two sets of BCDs; one is where
the processing is done on the basis of individual BCDs, and the other
has additional spatial and temporal filters applied that attempt to
remove instrumental signatures in an automated fashion.  (For a
description of the pipeline, see Gordon \etal\ 2005.)  We used the
filtered BCDs to construct mosaics for all of the targets at 70 \mum,
resampled to 4$\arcsec$ px$^{-1}$, about half the native pixel scale
of $9.85\arcsec\times10.06\arcsec$.

We extracted sources from the 70 \mum\ mosaics again using the APEX
1-frame portion of MOPEX.  For the sources that were detected, most of
the fluxes we report are from PRF-fitting; some bright source fluxes
are better determined using aperture photometry instead.  In those
cases, an aperture of 32$\arcsec$ and an aperture correction
(multiplicative factor) of 1.295 was used.  If no believable object
was seen by eye at the expected location, it was taken to be a
non-detection, and this aperture was laid down at the expected
location of the target, plus two other nearby locations $\sim1\arcmin$
north and south of the target position.  Based on these measurements,
an asssessment of the 1-$\sigma$ scatter per (native) pixel in nearby
background sky brightness was made over the aperture, and that scatter
was multiplied by 3 to obtain 3-$\sigma$ upper limits.  The same
aperture correction was used as for the aperture photometry of
detected objects.   

All but one of the targets was observed at 70 \mum.  CPD-64d120 was
not observed at 70 \mum\ because its expected photospheric flux was
far below the sensitivity that could be obtained within a reasonable
amount of integration time.  As can be seen in Table~3, 14 objects
were detected and 24 were not detected (3-$\sigma$ upper limits are in
Table~3).

The systematic uncertainty in the conversion of instrumental units to
calibrated flux units is estimated to be 5\% for default-scale
photometry by Gordon \etal\ (2008).  Gordon \etal\ are working with
PSF fitting; we have some PSF fitting and some aperture photometry. 
In addition, some of our targets are fainter than the ones used in
Gordon \etal, and some of our targets are observed in fine-scale
photometry mode.  To be conservative, then, we take the systematic
uncertainty to be 10\%.  Most of our objects are seen at
signal-to-noise ratios $>$ 10; our statistical error on detections is
much smaller than the systematic error of 10\% in most cases, so is
not reported.  In two cases, PZ Tel and $\epsilon$ Hya, the
statistical error (as determined with similar methdology to that for
the upper limits above) is comparable to the systematic error.  PZ Tel
is detected with a signal-to-noise ratio of $\sim$10 (uncertainty of 2
mJy on the 17.4 mJy detection in Table 3), and $\epsilon$ Hya is
detected with a signal-to-noise ratio of $\sim$3 (uncertainty of 4 mJy
on the 12.6 mJy detection in Table 3).

Several of our targets have serendipitously imaged detections in the
5$\arcmin\times$2.5$\arcmin$ field of view (see 
Appendix~\ref{sec:indobj}).  The density of extragalactic background
objects with brightness $\ge$ 15 mJy (the faintest 70 \mum\ detection
of a BPMG object achieved in this study) is 0.02 arcmin$^{-2}$ (Dole
\etal\ 2004). This leads to an expectation of 10 unrelated background
objects appearing in our data.  However, these are easily
distinguished from our target objects by their offset positions; the
probability is $< 1\%$ that a background object would be coincident
with any of our targets (see, e.g., Smith \etal\ 2006).

\subsubsection{160 \mum}

Twelve targets were observed at 160 \mum.  This subset of 12
targets was not selected uniformly for observation at 160 \mum.  For
the targets that were observed as part of program 102, those objects
expected to be brightest and seen at 70 \mum\ were selected for
observation at 160 \mum.  For the objects taken from other programs,
we have no way of reconstructing why these targets were selected for
observation.  

The MIPS data analysis tool (DAT) software (version 3.06; Gordon
\etal\ 2005) was used to calibrate the raw data ramp slopes, apply a
flat field correction, and mosaic the images in detector coordinates
at an image scale of 8$\arcsec$ pixel$^{-1}$ (half the native plate
scale of $15.96\arcsec\times18.04\arcsec$).  The data were
flux-calibrated using the standard  conversion factor of 1050 mJy
arcsec$^{-2}$ (flux unit)$^{-1}$, with about 12\% systematic
uncertainty (Stansberry \etal\ 2008).  For our oversampled image
mosaics, this is equivalent to 269 mJy DN$^{-1}$ in a pixel.

The MIPS 160 micron array suffers from a spectral leak that allows 
near-IR radiation to produce a ghost image adjacent to the true 160 
micron source image for stellar (roughly Rayleigh-Jeans) sources. The
leak is only bright enough to appear above the confusion noise  for
sources with $J\sim<$ 5.5 (MIPS Data Handbook V3.2).  Among our stars
observed at 160 \mum, three sources are brighter than this limit:
$\beta$ Pic, $\eta$ Tel, and AU Mic.  In the first source, the
circumstellar 160 \mum\ emission is considerably brighter than the
leak, so no effort was made to subtract off the leak.  In the latter
two sources, the leak  was subtracted using observations of a
photospheric standard, Achernar, from the Spitzer Archive (AORKEY
15572992) as a reference source.  The subtraction procedure is to
empirically determine the maximum normalization factor for the leak
reference source, such that its subtraction from the science target
does not produce noticable residuals below the background level.

To estimate upper limits to the 160 micron source flux
densities, we measured the rms background variation among four
7$\times$7 pixel apertures offset 64 arcsec along the detector
rows/columns from the  expected source position.  For this
aperture size, the 1-$\sigma$  equivalent noise was calculated
as 1/7th of the individual pixel rms,  assuming that the errors
combine in quadrature. This value was then converted to a
limiting flux density and corrected for the finite aperture size
using a multiplicative factor of 1.64 (measured from an STiny
Tim model PSF; Krist 2005).  Background cirrus emission can
cause variation in the  achieved sensitivity, with the
3-$\sigma$ upper limits ranging  over 27-77 mJy for our targets.

The five detections (and seven 3-$\sigma$ upper limits) are listed in
Table~3.  Most of our objects are seen at signal-to-noise ratios $>$8;
our statistical error on detections is much smaller in most cases than
the systematic error of 12\%, so is not reported. For $\eta$ Tel, the
detection has a signal-to-noise ratio of $\sim$4 (error of 16 mJy on
the 68 mJy reported in Table 3).

\subsection{Ancillary Data}

We consulted the literature for ancillary data on these objects
including spectral types, \ubvri\jhk\ and ground-based MIR magnitudes,
distances, \vsini, membership, etc.  References consulted for
literature values were the 2 Micron All-Sky Survey (2MASS;
Skrutskie \etal\ 2006),  NASA Star and Exoplanet Database (NStED; Ali
\etal\ 2005), the IRAS faint source catalog (FSC; Moshir \etal\ 1992)
and bright source catalog (BSC; Beichman \etal\ 1988), as well as
Zuckerman \& Song (2004), Song \etal\ (2003), Feigelson \etal\ (2006),
Chen \etal\ (2005b), Su \etal\ (2006), Kaisler \etal\ (2004), Zuckerman
\etal\ (2001), Mamajek \etal\ (2004), Zuckerman \etal\ (2001),
Plavchan \etal\ (2005), Jayawardhana \etal\ (2006), and Schneider
\etal\ (2006).

\subsection{SEDs and expected values}

Spectral energy distributions (SEDs) from $U$ band through 160 \mum,
created from the literature data plus our MIPS fluxes, for all of
these targets are portrayed in
Figures~\ref{fig:seds1}-\ref{fig:seds4}.  BPMG stars are in
Figures~\ref{fig:seds1}-\ref{fig:seds3}, and Tuc-Hor stars are in
Figure~\ref{fig:seds4}.  Note that the error bars are usually smaller
than the points; the points are hollow symbols, and the central
vertical bar is the corresponding error.

If available, spectral types as determined from spectra (not from
photometry) from the literature were used for each star. If no
spectrum-based type was available, the type determined from $V-K$
color as reported in Zuckerman \& Song (2004) was used; these types
appear in Table~1 as types without luminosity classes. Using
temperatures and gravities inferred from the spectral type, we
selected the closest grid point from the Kurucz-Lejeune model grid
(Lejeune \etal\ 1997, 1998).  This stellar model is shown in
Figures~\ref{fig:seds1}-\ref{fig:seds4} and is used to determine the
expected photospheric flux densities for the sample of stars at MIPS
wavelengths.  The models are normalized to the observed data at \ks. 
Interpolating models rather than selecting the nearest grid point does
not make any significant difference in the expected photospheric flux
density.  Using a single spectral type for unresolved binaries rather
than a hybrid of two spectral types also does not make any significant
difference in the expected photospheric flux density (see Trilling
\etal\ 2007). 

Because the spectral types are already well-known for most of these
stars, we did not wish to allow the spectral type to be a free
parameter in our fits.  However, we did wish to assess the goodness of
the fit.  Values of $\chi_{\nu}^2$ were calculated for the models in
Figures 1-4, and, as can be seen by eye in the Figures, all the fits
are quite good, even given occasional deviant optical points pulled
from the literature.  For most of the objects, typical values of
$\chi_{\nu}^2$ are $\sim$0.46 (\eg, typically $<$10\% chance that the
model is a bad fit). For the remaining objects, typically one optical
point is off (e.g., GJ 3322 A/B, see Figure 1), which distorts the
$\chi_{\nu}^2$; dropping those points brings the $\chi_{\nu}^2$ into
line with the rest of the objects.

The expected photospheric flux densities were linearly interpolated to
the MIPS wavelengths from the Kurucz-Lejeune model.   These estimated
photospheric flux densities are included in Table~3. If the assumed
spectral type is off by a subclass, over the entire range of types
considered in this paper, there is typically a $\lesssim$4\% change in
the calculated photospheric flux, comparable to the systematic
uncertainty in the measured 24 \mum\ fluxes.

\section{Results: Infrared Excesses in the BPMG}
\label{sec:findingdisks}

\subsection{Excesses at 24 \mum}
\label{sec:disks24}

\clearpage
\begin{figure*}[tbp]
\epsscale{1.0}
\plotone{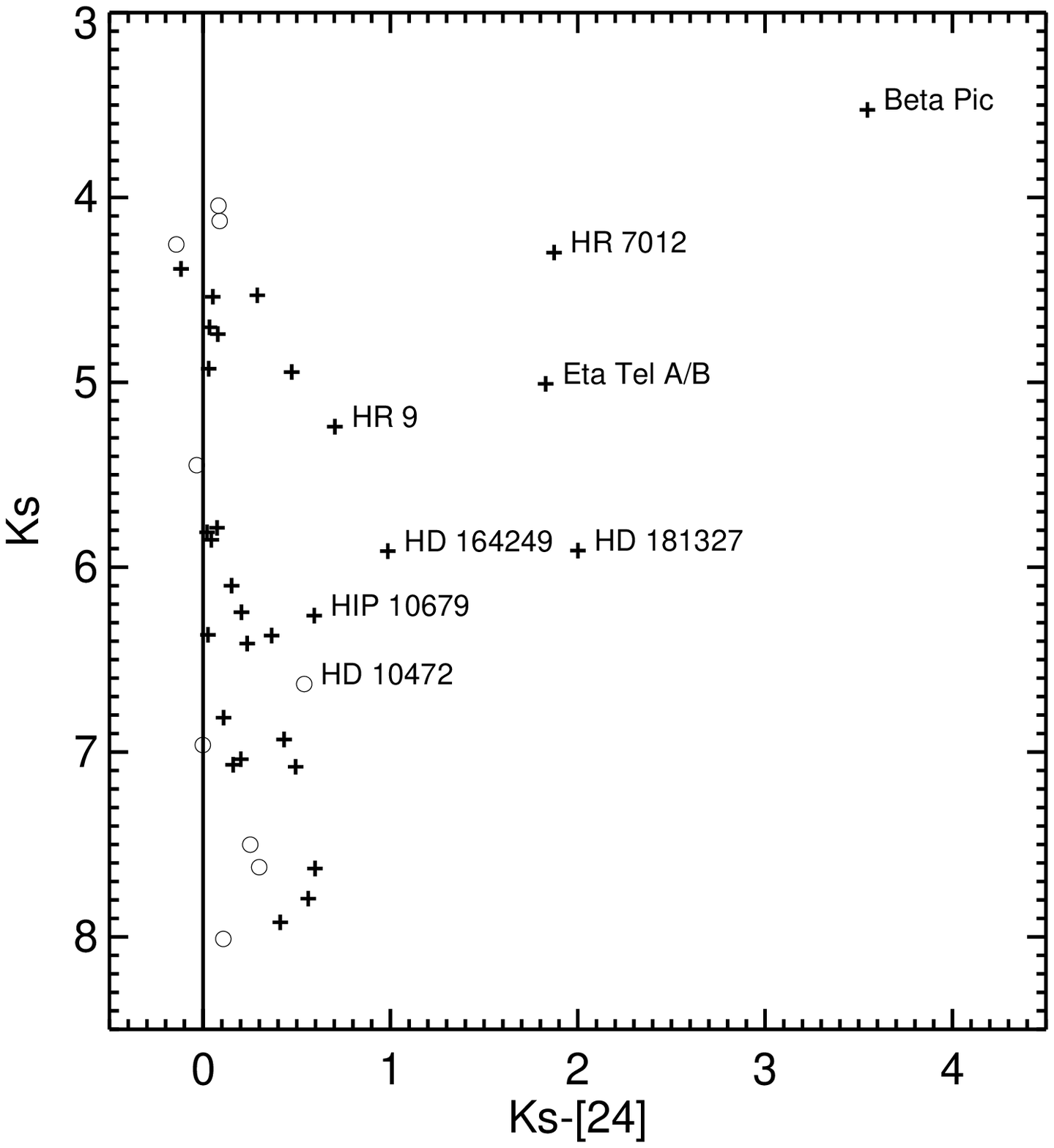}  
\caption{Plot of \ks\ vs.\ \ks$-$[24] for all of the objects
considered here.  Plus signs are objects from the BPMG; open circles
are objects from Tucanae-Horologium.  The  objects with excesses at 24
\mum, selected by a combination of techiniques, are indicated by name;
see text for discussion as to how the objects with excesses were
selected.}
\label{fig:kk24}
\end{figure*}

\begin{figure*}[tbp]
\epsscale{1.0}
\plotone{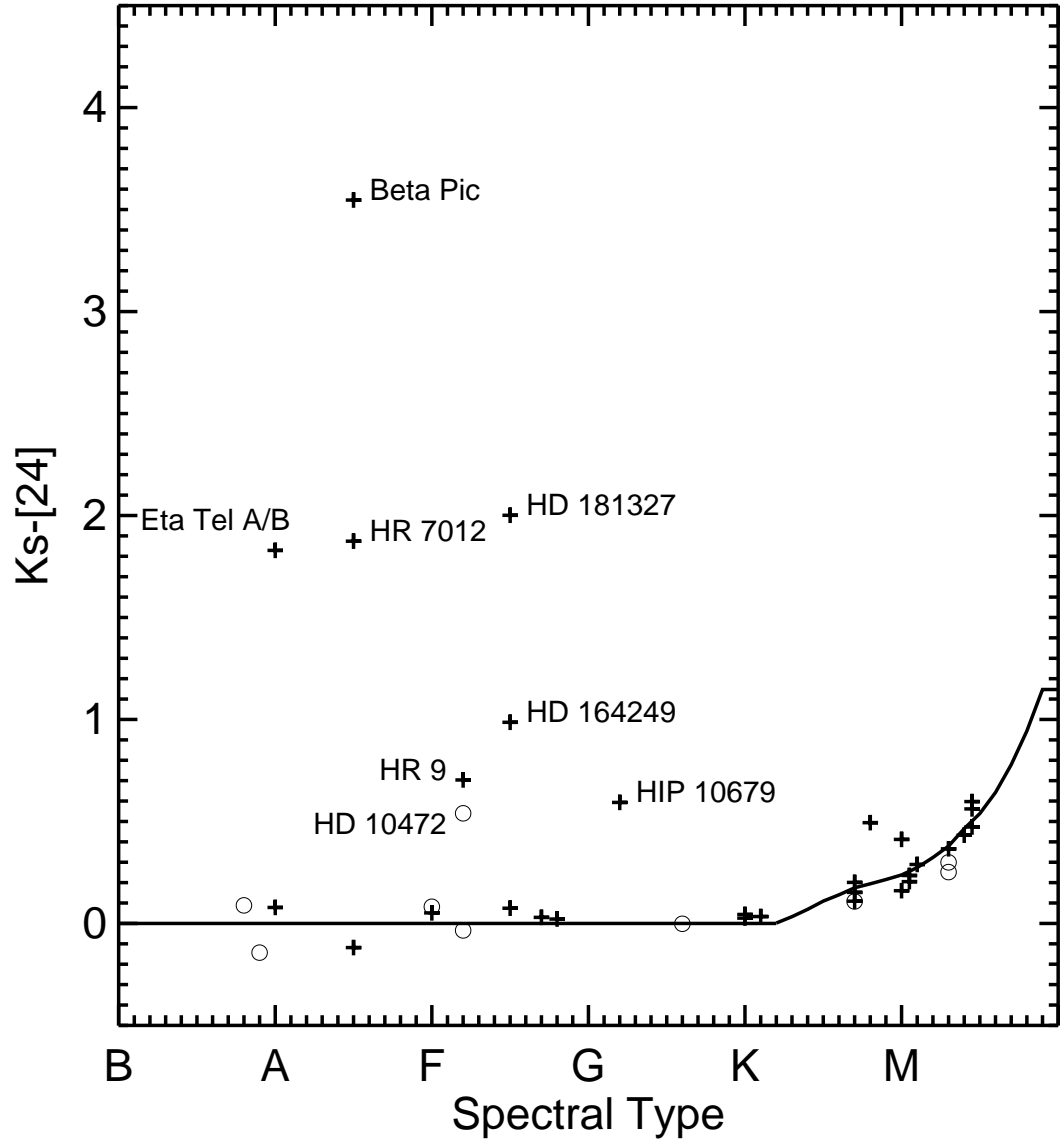} 
\caption{Plot of \ks$-$[24] vs.\ spectral type for all of the objects
considered here.  Plus signs are objects from the BPMG; open circles
are objects from Tucanae-Horologium.  The solid line indicates
expected photospheric color (Gautier \etal\ 2007). The  objects with
excesses at 24 \mum, selected by a combination of techiniques, are
indicated by name; see text for discussion as to how the objects with
excesses were selected.  }
\label{fig:kk24_spty}
\end{figure*}
\clearpage

There are a variety of methods in the literature for finding
circumstellar disks based on the 24 \mum\ excess. We will consider a
few slightly different methods here and establish our final sample of
7 objects (plus 1 more from Tuc-Hor) with excesses at 24 \mum. 

Figure~\ref{fig:kk24} is a color-magnitude diagram of \ks\ vs.\
\ks$-$[24].  In this figure, four stars stand out obviously as having
\ks$-$[24]$>$1.5: $\beta$ Pic, HR 7012, $\eta$ Tel, and HD 181327.  We
could declare these four stars as our only stars with excesses.
However, more subtle excesses are certainly present in the remaining
stars; the points do not scatter evenly around \ks$-$[24]=0.  Even
omitting the {\em five} stars with the largest \ks$-$[24], the mean
\ks$-$[24] color is 0.23, although there is a large standard
deviation; the 1$\sigma$ dispersion is 0.25.

To identify which of the remaining stars have excesses, we consider
the photospheric color.  For most spectral types, the
photospheric \ks$-$[24] color should be close to 0.  Gautier \etal\ (2007)
find for M stars that there is a dependence of \ks$-$[24] color with
\teff\ such that the latest types have a \ks$-$[24] color up to 1.5
for the coolest stars considered there (\teff$\sim$2000). 
Figure~\ref{fig:kk24_spty} shows the \ks$-$[24] color as a function of
spectral type for our sample here, along with the 
photospheric line from Gautier \etal\ (2007).  The non-zero color for
the latest types is readily apparent, and the tightness of the
correlation as a function of spectral type through the Ms clearly
follows the photospheres (see Gautier \etal\ 2007 for  more
discussion).  For types earlier than K0, we have another four stars
whose \ks$-$[24] colors are clearly distinct from 0: HD 164249 (F5),
HR 9 (F2), HIP 10679 (G2), and HD 10472 (F2).  These objects too are
therefore likely to possess excesses at 24 \mum.  

We clearly need to take into account expected photospheric flux to
assess the significance of the \ks$-$[24] excess, and for that we need
to depend on a model estimate of the photospheric flux. 
Bryden \etal\ (2006) consider nearby solar-type stars, calculating the
ratio of the measured to expected fluxes at 24 \mum.  They determined
infrared excess objects to be those with F$_{\rm meas}$/F$_{\rm
pred}>$1.2. Taking F$_{\rm meas}$/F$_{\rm pred}>$1.2 provides a
relatively conservative disk criterion at 24 \mum, in that it sets a
limit that is more than 3 times the systematic error of 4\%, also
providing ample room for the comparable $\sim$4\% uncertainty in the
calculation of F$_{\rm pred}$. In our sample, we can construct a
(sparse) histogram of F$_{\rm meas}$/F$_{\rm pred}$, and, as expected,
the histogram is sharply peaked around 1 with a break at 1.2 and a
long tail\footnote{To see the distribution of F$_{\rm meas}$/F$_{\rm
pred}$ in one dimension, see Figure~\ref{fig:ff24}.} extending out to
$\sim$27. The 1-$\sigma$ scatter in the points centered on F$_{\rm
meas}$/F$_{\rm pred}\sim$1 is 0.09.  The similar analysis in Bryden
\etal\ (2006) finds a 1-$\sigma$ scatter of 0.06.  The number we
obtain, 0.09, is an upper limit to the true error because we have
fewer stars than Bryden \etal\ and are using different methodology
(normalizing the star to \ks\ rather than fitting to the entire SED);
a few stars in our sample could inflate the error as a result of
small excesses or incorrect \ks\ magnitudes. On this basis, we believe
our results to be fundamentally consistent with those from Bryden
\etal\ (2006).  

The eight objects identified above have F$_{\rm meas}$/F$_{\rm
pred}>$1.2.  Note that $\beta$ Pic itself has a ratio of $\sim$27 (the
largest of the sample).  As shown in Figure~\ref{fig:kk24_spty},  HD
10472 has the smallest ratio, 1.6.  Assuming our scatter above of 0.09
as the worst-case-scenario, this lowest excess of our entire data set
is a 6-$\sigma$ excess.

There are two M stars in Figure~\ref{fig:kk24_spty} that could have
slight excesses, as their \ks$-$[24] are redder than other objects of
similar spectral type; they are AG Tri A and B. AG Tri A has a ratio
that is exactly 1.2.  AG Tri A will emerge in the next section as
having a 70 \mum\ excess, so it is quite possible that it has a small
24 \mum\ excess. AG Tri B has only a slightly larger \ks$-$[24]
than other stars plotted of similar spectral type, and has F$_{\rm
meas}$/F$_{\rm pred}$ at 24 microns of 0.98, well below our adopted
excess criterion.  

AU Mic is known to have a resolved disk at other wavelengths (e.g.,
Graham \etal\ 2007), so we investigated the evidence for an infrared
excess more closely.  The spectral type of this star is  usually taken
to be M1 (e.g., Graham \etal\ 2007, Houk 1982); it has 
\ks$-$[24]=0.29, which is comparable to the photospheric emission from
other stars of that spectral type from Gautier \etal\ (2007).  AU Mic
has F$_{\rm meas}$/F$_{\rm pred}$ at 24 \mum\ of 0.9, if anything
suggestive of a flux deficit at 24 microns.  If the spectral type of
AU Mic were incorrect, and its true spectral type was earlier, our
analysis method would yield a smaller value of F$_{\rm pred}$ and
hence a larger value of F$_{\rm meas}$/F$_{\rm pred}$.  In order to
yield F$_{\rm meas}$/F$_{\rm pred}>$ 1.2, however, the true spectral
type would have to be early K.  We have obtained our own high S/N,
echelle spectrum of AU Mic in order to constrain better its spectral
type (Stauffer \etal\ in preparation).  Based on the strength of the
TiO bandheads near 7050 \AA, we estimate a spectral type at least as
late as M1, and exclude a spectral type earlier than M0.  Therefore,
we believe our determination that AU Mic does not have an excess at 24
\mum\ is robust.  In support of this, we note that the IRAS 25 \mum\
flux is comparable to our 24 \mum\ flux, and that Chen \etal\ (2005)
also conclude that the star has no 24 \mum\ excess.

Formally adopting the Bryden \etal\ (2006) criterion, then, we find
that 8 out of the 39 stars or star systems in our entire sample have
24 \mum\ excesses. Out of the 30 stars in the BPMG for which we have
measurements, 7 have F$_{\rm meas}$/F$_{\rm pred}>$1.2.  Assuming that
the excesses are due to circumstellar disks, this implies a disk
fraction at 24 \mum\ of 23\%.

\subsection{Excesses at 70 \mum}
\label{sec:disks70}

\clearpage
\begin{figure*}[tbp]
\epsscale{1.0}
\plotone{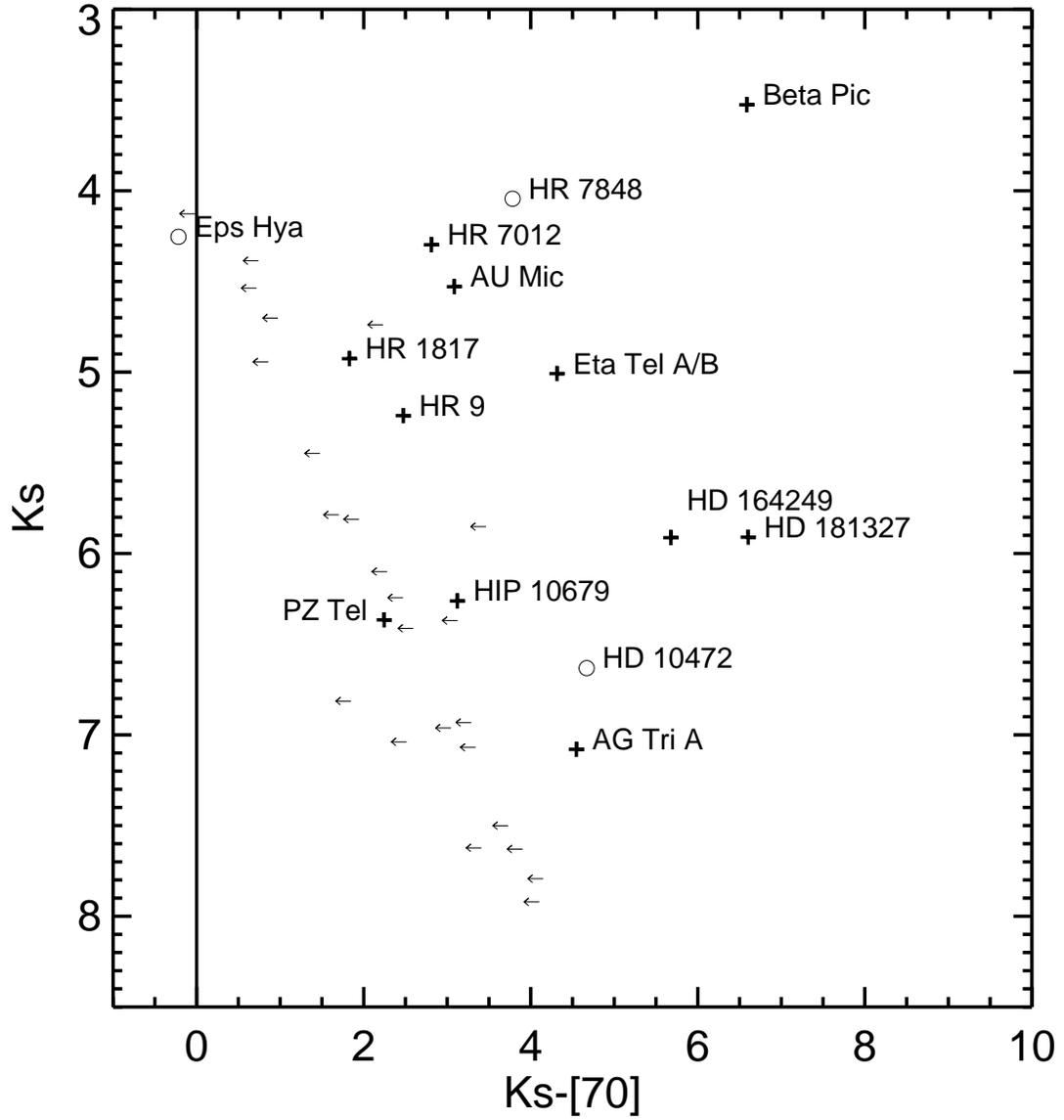}
\caption{Plot of \ks\ vs.\ \ks$-$[70] for all of the objects
considered here.  Plus signs are detected objects from the BPMG; open
circles are detected objects from Tucanae-Horologium. All other
objects (from both associations) are indicated as upper limits at 70
\mum.  All detections except $\epsilon$ Hya suggest excesses at 70
\mum.}
\label{fig:kk70}
\end{figure*}
\clearpage

Figure~\ref{fig:kk70} shows a plot similar to Figure~\ref{fig:kk24}
but for \ks$-$[70] colors.  Here, expected photospheric colors
are \ks$-$[70]$\sim$0.  Bryden \etal\ (2006) set a value of F$_{\rm
meas}$/F$_{\rm pred}\gtrsim$2 (the precise level is dependent on the
background level) to divide the disks from the non-disked stars.  The
overall scatter found for the Bryden \etal\ F$_{\rm meas}$/F$_{\rm
pred}$ for stars without excesses was 25\%, so the limit of  F$_{\rm
meas}$/F$_{\rm pred}\gtrsim$2 corresponds to 4$\sigma$.

In our data, just 14 stars are detected at 70 \mum\ (compared with 39
stars detected at 24 \mum).  We have many fewer detections than Bryden
\etal\ (2006), and even upon initial inspection of the SEDs or
Figure~\ref{fig:kk70}, just one object ($\epsilon$~Hya) seems to be a
likely photospheric detection.  In the 24 \mum\ section above, we were
able to examine the scatter of our measurements of F$_{\rm
meas}$/F$_{\rm pred}$ for photospheres in our sample;  there is no way
for us to repeat this analysis here at 70 \mum\ as a check on the
F$_{\rm meas}$/F$_{\rm pred}\gtrsim$2 excess cutoff. However, all of
the detections at 70 \mum\ are clear excesses, with the exception of
$\epsilon$~Hya.  Eps Hya has F$_{\rm meas}$/F$_{\rm pred}<$0.8, and
all the rest of the detections are F$_{\rm meas}$/F$_{\rm pred}>$4.9,
well beyond the Bryden \etal\ limit (20$\sigma$, assuming the 0.25
scatter), so we believe that the exact value for the criterion to
separate stars with excesses from those without is not critical. AG
Tri A, which was determined above to have an insignificant 24 \mum\
excess, has F$_{\rm meas}$/F$_{\rm pred}$=44 at 70 \mum. 

Fourteen of our larger sample of 38 stars or star systems are detected
at 70 \mum.  The sensitivity of the 70 \mum\ observations in this
sample varies considerably because of the range of exposure times used
in these observations and the cirrus background.  Out of the 30
members of the BPMG, 11 are detected, all of which have considerably
more than photospheric emission.  This represents a lower limit on the
BPMG excess fraction (at 70 \mum) of 37\%.

\subsection{Excesses at 160 \mum}
\label{sec:disks160}

None of the observations at 160 \mum\ are sensitive enough to detect
the expected photospheric flux densities, so all of the 160 \mum\
detections are suggestive of excesses.  Of the 12 BPMG stars with 160
\mum\ data, 5 are detected ($\beta$ Pic, HD 164249, $\eta$ Tel, HD
181327, and AU Mic).  Because the sample of stars selected for
observation at 160 \mum\ is biased towards those with disks, we cannot
infer a limit on the excess fraction (at 160 \mum).

All stars detected at 160 \mum\ are also detected at 70 and 24 \mum,
and almost all of the stars with 160 \mum\ excesses also have excesses
at the other two MIPS wavelengths.  The sole exception is AU Mic,
which has a clear excess at 70 but not at 24 \mum.  Based on the
blackbody fits (see below), in no case does the 160 \mum\ detection
suggest a cold component of dust that is not seen at the shorter MIPS
wavelength(s). (For a discussion of how much cold dust could be
included within the uncertainty of the 160 \mum\ detections that is
not already accounted for with the component seen at 24 and 70 \mum,
please see Gautier \etal\ 2007.)

\subsection{Comparison with IRAS}

Of our 39 targets observed at 24 \mum\ from both the BPMG and Tuc-Hor,
19 appear in the IRAS FSC or BSC with either detections or upper
limits (plus 3 more included with a nearby association member by the
IRAS beam).  Discussion of individual objects is in
Appendix~\ref{sec:indobj}, including those objects where MIPS
observations have resolved source confusion (or background
contamination) found in the large-beam IRAS measurements.  

In summary, the MIPS observations confirm five excesses discovered by
IRAS. In five more cases (four of which have excesses at 70 \mum),
MIPS provides a detection near the IRAS limit.  For the remaining 9
systems, MIPS establishes a new much more stringent upper limit on any
excess that may be present; at least 3 of those previously appeared to
have an excess based solely on IRAS results.  There are three new
excesses without any prior IRAS detections or limits.  For individual
source assessment, the SEDs for each object (including IRAS detections
and limits) appear in Figures~\ref{fig:seds1}-\ref{fig:seds4}, and
discussion of specific cases appears in the Appendix.

\section{Disk Properties}
\label{sec:models}

We note here for completeness the following items.    Simply having
$L_{\rm dust}/L_{*} < 1$ does not assure that the disk is really a
debris disk, which by definition requires a second generation of dust
and gas depletion; however, values of $L_{\rm dust}/L_{*} \sim
10^{-3}$ are likely debris disks.  
Spitzer/MIPS observations constrain the presence of dust in these
systems, but say nothing about any gas or grains much larger than the
wavelength of observation.  From this point forward, we have assumed
that any excess infrared emission that we observe above the 
photosphere is due entirely to a dusty circumstellar disk.  Until
observations at any wavelength resolve the disk, this remains an
assumption.

\clearpage
\begin{deluxetable}{lrrrrrrr}
\tablecaption{Model Results: Disk Properties}
\label{tab:results2}
\tablewidth{0pt}
\tabletypesize{\scriptsize}
\tablecolumns{8}
\tablehead{
&  \multicolumn{4}{c}{simple blackbody models} & 
\multicolumn{3}{c}{more complex models} \\
\colhead{object} & \colhead{BB T}
& \colhead{$L_{\rm dust}/L_{\rm *}$ } &
\colhead{min. $R_{\rm dust}$ } & 
\colhead{min. $M_{\rm dust}$ } &
\colhead{$M_d$} & \colhead{$R_i$} & \colhead{$R_o$} \\
 & \colhead{(K)}
& \colhead{ ($\times10^{-5}$)} &
\colhead{(AU)} & 
\colhead{($M_{\rm{moon}}$)} &
\colhead{($M_{\rm{moon}}$)} & \colhead{(AU)} & \colhead{(AU)}}
\startdata
\cutinhead{Disks detected at more than one wavelength}
           HR 9&  120 & 10      & 10 & 0.0004   & 0.25 & 35 & 200\\ 
      HIP 10679&  100 & 80      & 20 & 0.01 & 0.4 & 35 & 200 \\
       AG Tri A\tablenotemark{a}&  65  & 79      & 10 & 0.003  & \ldots\tablenotemark{d} & \ldots & \ldots\\ 
$\beta$ Pic\tablenotemark{b}&  130 & 180 & 10 & 0.012& \ldots & \ldots & \ldots\\
      HD 164249&  78  & 59      & 20 & 0.01 & \ldots\tablenotemark{d} & \ldots & \ldots\\  
        HR 7012&  310 & 90      & 2  & 0.0002   & 0.05 & 5 & 200\\  
 $\eta$ Tel A/B&  140 & 24      & 20 & 0.0027  & 0.8 & 70 & 200\\
      HD 181327&  75  & 250     & 20 & 0.06 & 10 & 68\tablenotemark{c} & 104\tablenotemark{c}\\ 
         AU Mic&  50  & 23      & 8  & 0.0005 & 1 & 35 & 200\\ 
      HD 10472 (Tuc-Hor) &  70  & 67      & 30 & 0.02 & 30 & 400 & 700\\ 
\cutinhead{Disks detected only at 70 \mum\ }
        HR 1817&  (41) & $>$3.0 &  (60) &  (0.004) & 0.3 & 100 & 200\\ 
         PZ Tel&  (41) & $>$7.3 &  (50) &  (0.006) & 0.3 & 35  & 200\\  
        HR 7848 (Tuc-Hor) &  (41) & $>$13  & (100) & (0.07) & 5   & 250 & 400 \\ 
\enddata
\tablenotetext{a}{Since the F$_{\rm meas}$/F$_{\rm pred}$ at 24 \mum\
for this star was right at 1.2, we attempted modelling of this star
including the observed flux density at 24 \mum.}
\tablenotetext{b}{A simple disk fit was made for Beta Pic for
self-consistency with the rest of the sample; this object is resolved
at MIPS wavelengths and the disk is better characterized using other
methods.}
\tablenotetext{b}{Inner and outer radii are fixed at the values
reported by Schneider \etal\ (2006).}
\tablenotetext{c}{No fit possible; see text for discussion. }
\end{deluxetable}
\clearpage

\subsection{Blackbody fits}

For those 13 objects that we find to have excesses at any MIPS
wavelength, Figures~\ref{fig:seds1} through \ref{fig:seds4} show a fit
to the star+disk SED.  The excesses are modelled as simple blackbodies,
which we use for an initial simple characterization of the disks, akin
to an assumption of a single-temperature thin ring of dust.  

In three cases (HR 1817, PZ Tel, HR 7848), we have a single data point
at 70 \mum\ that describes the disk excess.  For these objects, we
follow the example set by Bryden \etal\ (2006) and simply set the peak
of the blackbody to be at 70 \mum\ (41 K for $\lambda F_{\lambda}$). 
In ten cases, we have more data (detections and limits) that describe
the disk; for these stars, we have found the best-fit blackbody by
$\chi^2$ minimization analysis, allowing the best-fit blackbody to run
through the upper limits where available. The temperatures
corresponding to those fits can be found in Table~4.   Note that in
the case of AG Tri A, the F$_{\rm meas}$/F$_{\rm pred}$ at 24 \mum\ is
1.2, so a small excess at 24 \mum\ cannot be ruled out; we modelled
this star including this potentially small excess at 24 \mum, so it is
effectively treated as a star with more than one disk detection.

Since a blackbody has two free parameters, disks with two data points
describing the disk are fit perfectly by a blackbody, and this can
clearly be seen in Figures~\ref{fig:seds1} -- \ref{fig:seds4}.  We do
not expect a simple blackbody to be a good fit to disks with three
data points, because in reality there is wavelength-dependent grain
emissivity for small grains that is not accounted for in a simple
blackbody, and there is likely to be dust with a range of
temperatures.  Clearly better models than a simple blackbody are
needed to characterize the disks (see below).   Nonetheless, as can be
seen in Figures~\ref{fig:seds1} -- \ref{fig:seds4}, the fits are
acceptable for even the four objects ($\beta$ Pic, HD 164249, $\eta$
Tel, and HD 181327) with excesses at all three MIPS bands, although,
not surprisingly, many are not within 1$\sigma$ of the data points. 
The fit for HD 164249 is the most discrepant, running below the 70
\mum\ point (333 mJy predicted by the model, compared to 624 mJy
observed) but above the 160 \mum\ point (170 mJy predicted vs.\ 104
mJy observed). In this case in particular, the dust distribution may
well be impossible to characterize with a single temperature simple
blackbody, even with grain emissivity included -- for example, there
may be a range of particle sizes and a large distribution of orbital
radii.  Indeed, spectral features have been resolved from disks around
$\beta$ Pic, HR 7012, and $\eta$ Tel (Chen \etal\ 2006, 2007).
Nevertheless, for completeness and self-consistency within the sample,
we list the numbers obtained via the simple blackbody fit in
Table~4.  

The hottest dust found in the sample is $\sim$300 K for HR
7012.  AU Mic's disk, which is resolved by other instruments
(e.g., Kalas \etal\ 2004, Krist \etal\ 2005) though not by MIPS
(Chen \etal\ 2005), is fit by the coldest dust of any of these
objects (especially among those with 160 \mum\ detections) at
$\sim$50~K, which is consistent with a disk excess at 70 and 160
but not 24 \mum.  

Although we also fit $\beta$ Pic, AU Mic, and HD 181327 with single
blackbodies  for self-consistency within the sample and for comparison
here, we note that these objects are resolved at other wavelengths --
$\beta$ Pic is resolved even at MIPS wavelengths (Su \etal\ 2004) and
is known to not be a single-temperature narrow ring --  so their disks
are better characterized using other methods that take into account
that spatial information.

\subsection{Fractional IR excess}

Since we have a wide range of spectral types represented in this
association, we would like to use a measurement of the disk
luminosity that attempts to compensate for the central star's
luminosity.  We used the fits described above to derive a value
for the fractional disk luminosity, $L_{\rm dust}/L_{*}$; these
values appear in Table~4.  To determine $L_{\rm dust}$ for stars
which have an excess described by more than one detection, we
integrate under the disk model fit, having subtracted off the
photospheric contribution.  In order to determine the $L_{\rm
dust}/L_{*}$ value for stars whose excesses are only observed at
70 \mum, we follow Bryden \etal\ (2006; equation 3), determining
the minimum $L_{\rm dust}/L_{*}$ by assuming that the blackbody
continuum peaks at 70 \mum.  

The $L_{\rm dust}/L_{*}$ values that appear in Table~4 for disks
detected in more than one wavelength range from $10^{-4}$ to
2.5$\times 10^{-3}$, with a median value of $7.9\times 10^{-4}$.

\subsection{Minimum Radius and Minimum Mass}
\label{sec:lowmodels}

Assuming that the grains composing the disks are in thermal
equilibrium, we can follow a similar analysis as found in Low
\etal\ (2005) or Smith \etal\ (2006) to determine a minimum
radius and minimum mass of the disk.  We assume blackbody dust
grains in thermal equilibrium with the stellar radiation field,
and contrain the inner radius of the disk along
with a minimum mass of the disk.  Following Low \etal\ (2005),
we use the relationship from Chen \& Jura (2001).  We assume the
same values for average grain size (2.8 \mum) and density (2.5 g
cm$^{-3}$) adopted there (and in Low \etal\ 2005 and Smith
\etal\ 2006), despite the fact that these parameters, having
been derived for $\zeta$ Lep (an A3 star), may be more
appropriate for much more massive stars than we have here on
average (see additional discussion below).  Values of minimum
radius and minimum mass so calculated appear in Table~4.  For
disks detected in more than one wavelength, the minimum radius
ranges from 2 to 30 AU, and the minimum mass ranges from
$\sim$0.0002 to $\sim$0.06 $M_{\rm{moon}}$.

\subsection{Literature Models}

Spitzer Infrared Spectrograph (IRS; Houck \etal\ 2004) observations of
HD 181327, HR 7012, and $\eta$ Tel were discussed and modelled in Chen
\etal\ (2006).  The IRS observations extend to 33 \mum.  The MIPS-24
\mum\ flux densities are consistent with the IRS spectra; since the
Chen \etal\ (2006) models were designed to fit IRS spectra between 4
and 33 \mum, of course the models are also, by construction,
consistent with our MIPS-24 \mum\ flux densities.  In all three cases,
these models can be extended past 33 \mum\ to predict flux densities
at 70 and 160 \mum, and they are found to be in very good agreement
with the observed flux densities.  

For HD 181327, the IRS spectrum is featureless and Chen \etal\ model
the excess as a simple blackbody, making it straightforward to compare
their model parameters to ours.   The blackbody temperature from Chen
\etal\ is 81K; our blackbody  temperature is 75K, which we consider to
be identical to within the errors.  The $L_{\rm dust}/L_{*}$ reported
by Chen \etal\ is 3.1$\times10^{-3}$, to be compared with
2.5$\times10^{-3}$ derived here.  The minimum mass of the disk is
4$\times10^{24}$ grams in this paper and 1$\times10^{24}$ grams in
Chen \etal.  The Chen \etal\ model predicts a 70 \mum\ flux density of
1.2 Jy (20\% different than observations) and a 160 \mum\ flux density
of 0.62 Jy (3\% different than observations).   

For the other two stars (HR 7012 and $\eta$ Tel), Chen \etal\ found
features in the IRS spectra and constructed much more detailed, 
multi-component models (with various mineral species and a range of
grain sizes, etc.), making comparison to parameters derived from our
single-component blackbody fits relatively unilluminating.    However,
in order to match the overall structure of the IRS spectra found near
$\sim$30 \mum, Chen \etal\ required a cooler component, up to two
blackbodies of different temperatures and total solid angles.  The
assumptions of the models are sufficiently different from ours as to
make simple comparisons difficult.  These differences simply
illustrate the latitude even relatively detailed models have in
fitting the existing data, given the large number of parameters that
can be adjusted. The one somewhat meaningful comparison is of the
blackbody temperatures to fit the longest wavelength flux densities.  
For HR 7012, Chen \etal\ adopted a blackbody temperature of 200 K,
versus 310 K for our models; for $\eta$ Tel, Chen \etal\ adopted two
blackbodies, one of 115 K and the other of 370 K (however, the total
solid angle of the 115 K component was much larger), versus our 140 K
single blackbody.   For HR 7012, the predicted flux densities are 0.27
and 0.072 Jy at 70 and 160 \mum, respectively; at 70 \mum, the
observed flux density is 35\% different from the model, and it was not
observed at 160 \mum.   For $\eta$ Tel, the Chen \etal\ predicted flux
densities are 0.39 and 0.14 Jy, and our observed values are just 5 and
8\% different from the model. 


\subsection{New Models}

\clearpage
\begin{figure*}[tbp]
\epsscale{0.8}
\plotone{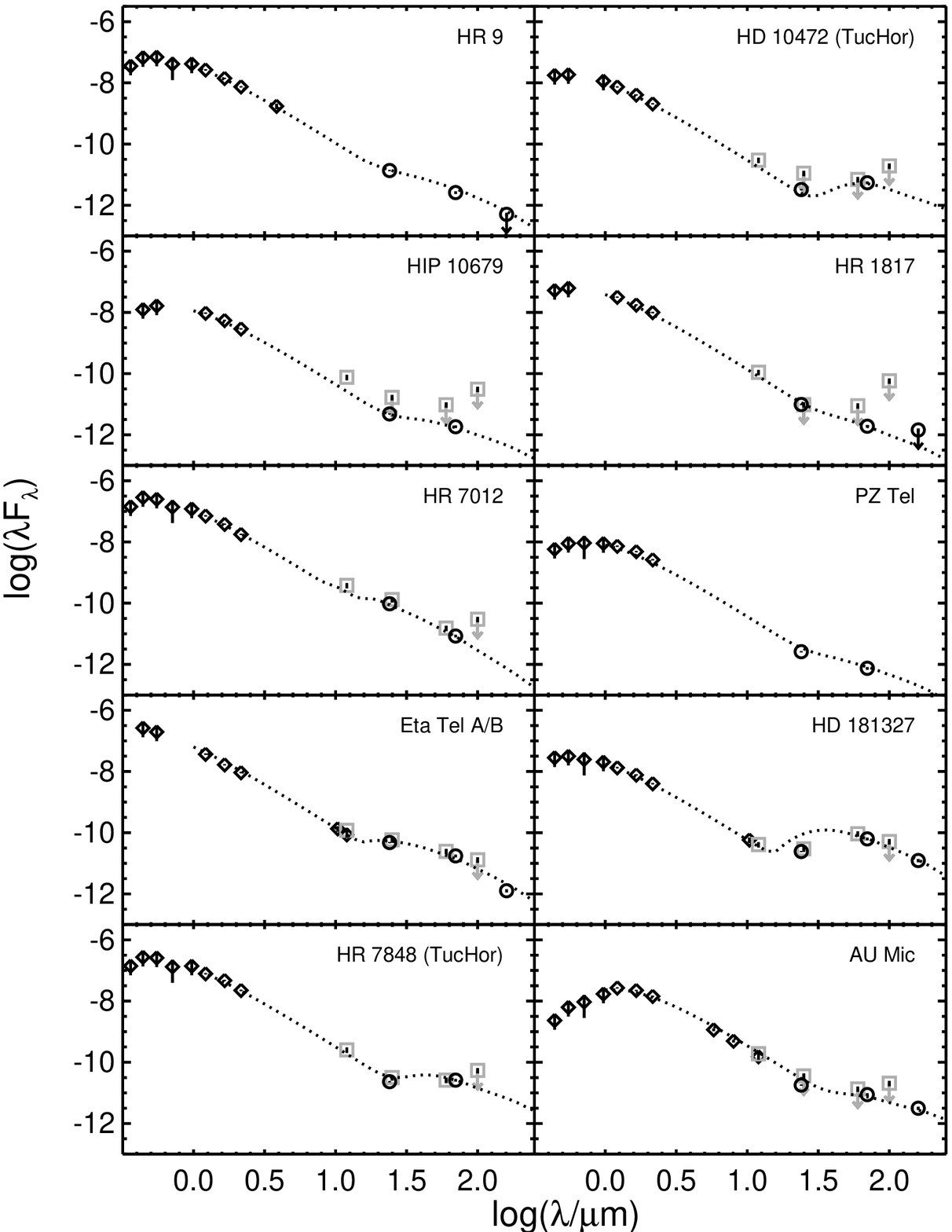}
\caption{Star+disk models of the ten stars considered for more
sophisticated modeling; see text for discussion as to how they were
selected and the details of the modelling.  Notation is as in
Figures~\ref{fig:seds1}-\ref{fig:seds4}: the
$x$-axis plots log of the wavelength in microns, and the $y$ axis
plots log($\lambda F_{\lambda}$) in cgs units (ergs s$^{-1}$
cm$^{-2}$).  Points gleaned from the literature are diamonds, boxes
are detections or upper limits from IRAS, and circles are new MIPS
points.  Downward-pointing arrows indicate upper limits.}
\label{fig:vgm1}
\end{figure*}

\begin{figure*}[tbp]
\epsscale{0.8}
\plotone{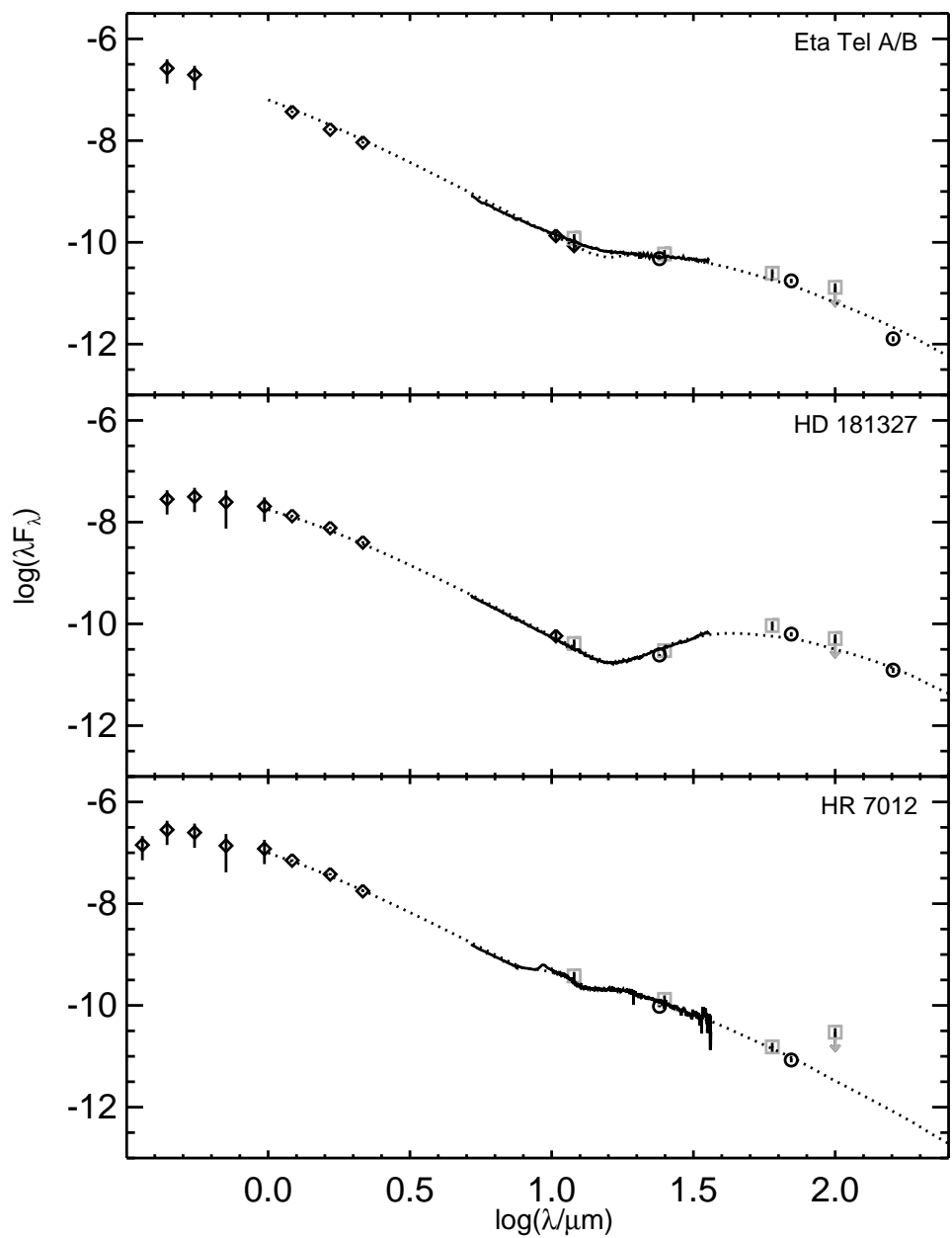}
\caption{Star+disk models of the three stars with IRS spectra
considered for unconstrained modeling; see text for details of the
modelling.  Notation is as in Figures~\ref{fig:seds1}-\ref{fig:seds4};
the $x$-axis plots log of the wavelength in microns, and the $y$ axis
plots log($\lambda F_{\lambda}$) in cgs units (ergs s$^{-1}$
cm$^{-2}$).  Points gleaned from the literature are diamonds, boxes
are detections or upper limits from IRAS, circles are new MIPS points,
and the solid line is the IRS spectra from Chen \etal\ (2006). 
Downward-pointing arrows indicate upper limits.}
\label{fig:vgm2}
\end{figure*}
\clearpage

Thirteen of the 40 targets have flux excesses above photospheric
levels in at least one of the MIPS bands. Of these, $\beta$ Pic
itself, has been studied extensively in the literature (most recently
Chen \etal\ 2007), and we consider it no further here. We have fit the
data points as portrayed in Figures 1-4 for the remaining twelve
systems using continuum spectra computed in each case for an
axisymmetric and optically thin disk of astronomical silicate grains
in radiative equilbrium with the stellar field. The models are
described further in Mannings \etal\ (in prep); below, we summarize
the characteristics of the models. 

\subsubsection{Model Description}

We assume grain radii distributed as a power law from 0.001 microns to
1 mm. The index of the {\it continuous} power law distribution in
grain size is here fixed at $-$2.5, leading directly from the index of
$-$3.5 for the number of grains {\it per unit size interval} 
described in the clasic study of interstellar grains by Mathis, Rumpl
\& Norsieck (1977).  Optical constants are taken from Draine (2007)
for the smaller grains. We compute absorption efficiencies for the
larger grains by modifying the Mie code developed by Bohren and
Huffman (1983). We then distribute the grains across a disk geometry
assuming a surface density viewed normal to the disk plane that falls
off as a power law from an inner disk radius $R_i$ to an outer radius,
$R_o$.  (See  Sylvester \& Skinner 1996 for similar modelling of
debris disks.) The power law index for the radial density distribution
is held at the typical value of $-1.5$ assumed for circumstellar disks
(\eg, Kenyon \& Bromley 2002). The disk inclination angle is
irrelevant for optically thin emission, as is the (likely) non-zero
opening angle of the disk as viewed from the star. The remaining disk
parameter is simply the total mass of grains, $M_d$. To limit the
number of free parameters (since in several cases we have but one
point defining the disk), we fix all quantities with the exception of
$R_i$, $R_o$, and $M_d$.  These three parameters dominate in different
wavelength regimes, so we are able to hone in on a unique fit despite
the sparseness of the data. To first order, the value of $R_i$
establishes, for a fixed range of grain sizes, as in this model, the
wavelength at which the disk spectrum exhibits a peak, while  $M_d$
determines the luminosity of the disk. The spectrum is relatively
insensitive to $R_o$. Increasing the value of $R_o$ relative to $R_i$
is akin to spreading the grains out to greater distances from the star
but, since the radial density falls as a power law, the effect on the
total spectrum is marginal. It can be perceived as a gentle softening
of the ratio of the total flux emitted by warm grains (inner disk) and
cool grains (outer disk). We show our model SEDs in 
Figure~\ref{fig:vgm1}. 


\subsubsection{Model Results and Comparison}

The best fit values are listed in Table 4 for $M_d$, $R_i$ and $R_o$,
as derived using these models and the optical+near IR+MIPS data that
appear in Figures 1-4. Disk masses range from about 0.05 to 30 Lunar
masses. Disk inner radii take values from 5 to 400 AU, and outer radii
range from about 100 to 700 AU.  The median fractional difference
between the model and the observations at 24 \mum\ is $-$0.12,
including the value from HD 181327, which is the most discrepant at 24
\mum\ (see Figure~\ref{fig:vgm1}, and discussion below).  The closest
fit is HIP 10679, where the model matches the observations to 3\%. 
Given that the systematic uncertainty of our 24 \mum\ observations is
4\%, the model is then typically $\sim3\times$ off at 24 \mum.  At
both 70 and 160 \mum, the median fractional difference between the
model and observations is just 0.04, well within the systematic
uncertainty at either band.

The simpler models calculated following Low \etal\ (2005) in
Section~\ref{sec:lowmodels} above (hereafter abbreviated as
``Model 1''), not surprisingly, produce much different values of
disk masses and radii than those calculated here.  The models
from Mannings \etal\ (hereafter abbreviated ``Model 2") are more
complex; both   Models 1 and 2 are physically valid within the
limitations of their own set of assumptions, which we now
discuss.

Model 1, in order to calculate the minimum disk mass and inner minimum
disk radius, must make simple assumptions about the grain size  (2.8
\mum) and density (2.5 g cm$^{-3}$), and assume that the grains
radiate as blackbodies.  These assumptions trace back to Chen \& Jura
(2001), who studied an A3 star, $\zeta$ Lep; they took 2.8 \mum\ for
grain size because grains smaller than this would be ejected from the
system due to radiation pressure.  This is {\em not} a universally
valid assumption for these BPMG stars (or for that matter for the TWA
stars from Low \etal\ 2005), because there are much cooler M stars
included in both BPMG and TWA.  But, such calculations nonetheless
serve to provide a rough comparison between star-disk systems across
papers and associations.  

Model 2 obtains such different results for disk masses and sizes for a
variety of reasons, all traced back to grain size and location
assumptions. Model 2 assumes that each disk is a power-law mixture of
grain sizes (from ISM size to 1 mm), and that the mixture is spread
out across the disk (not in a thin ring).  Most of the grains are a
factor of 3000 smaller in radius than that assumed in Model 1, and the
grain emission is not blackbody.  Small non-blackbody grains tend to
be hotter than larger (\eg, blackbody) grains at the same distance
from a star, so the small grains must be further out to get lower
temperatures and, therefore, similar fluxes. That in part accounts for
the Model 2 disk inner radii being larger than those of Model 1. 
(Moreover, the radii from Model 1 are artificially reduced by the
assumption that the particles radiate like blackbodies at the
temperatures or wavelengths of interest, which is almost certainly not
the case as even 3 \mum\ particles are small compared to the relevant
wavelengths.) Because Model 2 has larger disk radii, a much larger
amount of dust area is needed to subtend a given solid angle to absorb
the stellar light and match the observations.  The Model 2 disk
masses are larger than those of Model 1 for two reasons. First,
because the best-fit inner disk radii are larger than the Model 1
values, a greater amount of integrated grain surface area is needed in
Model 2 to subtend a similar total solid angle, as viewed from the
star, to that for Model 1. Second, due to the power-law distribution
in grain sizes, a significant amount of disk mass is locked up in the
large end of the size range, while the absorption and re-emissison of
starlight is dominated by grains at the small end. The small grains
contribute negligibly to the disk mass, but they dominate the
radiative transfer and, therefore, the output spectrum.


\subsubsection{Notes on models of specific sources}

For HD 181327, the inner and outer radii were fixed at the values
reported by Schneider \etal\ (2006), despite the fact that those
parameters were obtained from wavelengths shorter than 24 \mum.  Only
the mass was left as a free parameter in our model fit.  This (plus
the other constraints imposed) explains why the predicted model flux
density at 24 \mum\ is so different than the observed flux density
(see Figure~\ref{fig:vgm1}, and below).  

Two of the twelve sources with flux excesses cannot be fit with model
disk spectra: AG Tri A and HD 164249. The MIPS detections for these
latter targets could include background sources that cannot be
distinguished from the target stars, but as we argue above, this is
relatively unlikely, $<$1\%.  It is more likely that some of the fixed
parameters need to vary, and that measurements are needed at other
wavelengths to constrain the models.  Both of these objects are also
not particularly well-fit by the simple blackbodies above. HD 164249
was called out as a particularly poor fit above; with the more
sophisticated modelling (given the constrained parameters above), the
24 \mum\ excess can be fit, but the 70 \mum\ model is well below the
observed flux.  AG Tri A's simple blackbody fit above runs through the
upper limit at 160 \mum, and if the true flux of the system is really
much lower, the simple fit will not work either.  

\subsubsection{Testing the simple models by including IRS data}

Three stars have IRS spectra as noted above and as reported in Chen
\etal\ (2006) -- $\eta$ Tel, HD 181327, and HR 7012.  (Additional IRS
spectra for several more BPMG stars exist in the Spitzer Archive, but
analyzing those data is beyond the scope of this paper.)  As a simple
way of assessing the limitations of the simple models performed above
that primarily rely on the MIPS data in the mid- and far-IR, for
$\eta$ Tel, HD 181327, and HR 7012, we included the IRS data and then
attemped an unconstrained Mannings \etal\ model fit, e.g., letting all
of the parameters vary.  Plots of these fits (including the IRS data
from Chen \etal\ 2006) appear in Figure~\ref{fig:vgm2}.

For $\eta$ Tel, the constrained model fit above slightly underpredicts
the 70 \mum\ flux density (by 17\%) while slightly overpredicting the
160 \mum\ flux density (by 10\%). In order to fit the IRS data as
well, the best model fit now brings the inner radius in from 70 to 30
AU, and the disk mass from 0.8 to 0.3 $M_{\rm moon}$.  

For HD 181327, the constrained model fit above predicts a 
higher 24 \mum\ flux than is observed.  In order to allow the
model to fit the IRS+MIPS data together, but still leave the
inner disk radius constrained to that reported by Schneider
\etal\ (2006), we increased the minimum grain size from 0.001
\mum\ to 1 \mum, so the grains are distributed as a power law
from 1 \mum\ to 1 mm.  The model matches the IRS spectrum very
well, eliminating the discrepancy at 24 \mum, but slightly
underpredicting (by 17\%) the 70 \mum\ flux while slightly
overpredicting (by 10\%) the 160 \mum\ flux.  The disk mass
increases from the 9 $M_{\rm moon}$ reported above to 11 $M_{\rm
moon}$. 

Finally, for HR 7012, the best-fit disk mass is identical to the fit
as reported above (0.05 $M_{\rm moon}$) and the inner disk radius
changes from 5 to 3.5 AU, not a significant change.  The model
replicates well the emission features observed near 10 and 20 \mum, so
the grains in this disk could be silicate or have a large silicate
component, as reported by Chen \etal\ (2006).

\clearpage
\begin{figure*}[tbp]
\epsscale{0.8}
\plotone{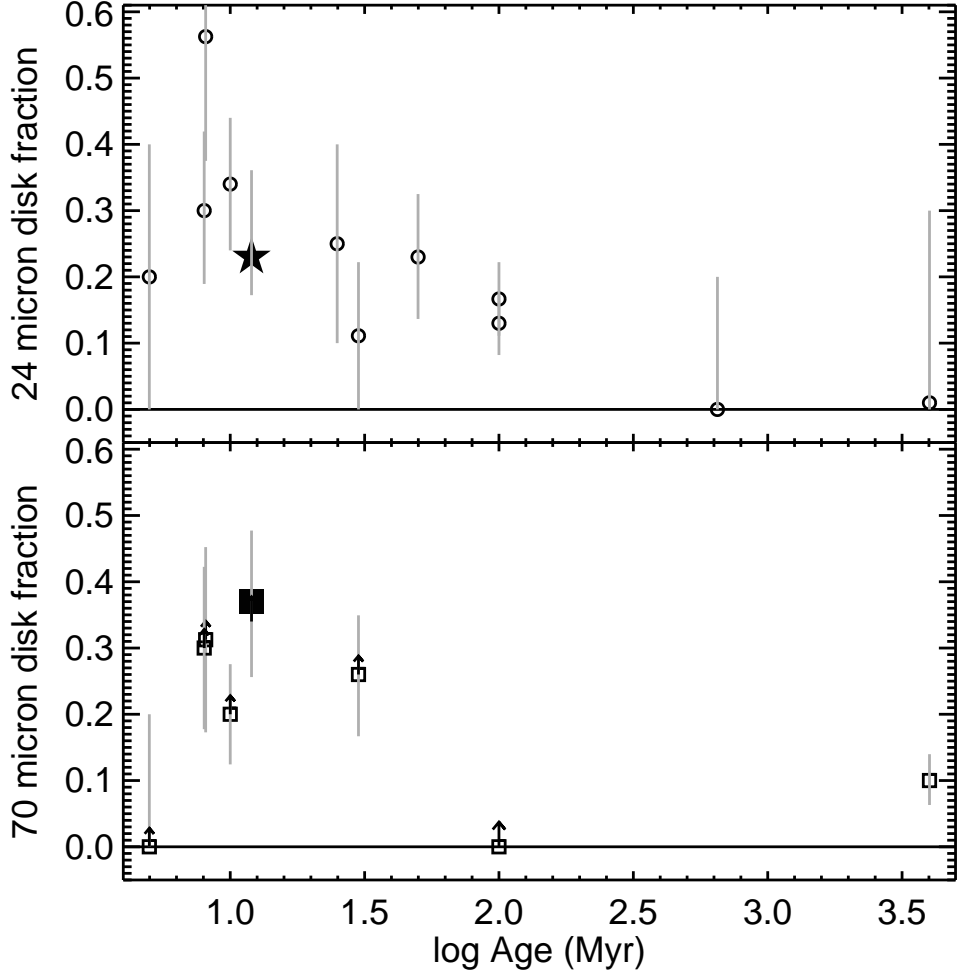}
\caption{Evolution of disk fraction with time: the top panel is the 24
\mum\ disk fraction, and the bottom panel is the 70 \mum\ disk
fraction.  Values from the literature (see Table~5) are compared with
our values for the BPMG. Literature 24 \mum\ points are circles, BPMG
24 \mum\ point is a large solid 5-pointed star, literature 70 \mum\
points are boxes, and the BPMG 70 \mum\ point is a large solid box.
Grey vertical lines are the errors calculated from Poisson (counting)
statistics.  Our points are consistent with the disk fractions for
similarly-aged clusters and associations found in the literature.}
\label{fig:evol}
\end{figure*}

\begin{figure*}[tbp]
\epsscale{1.0}
\plotone{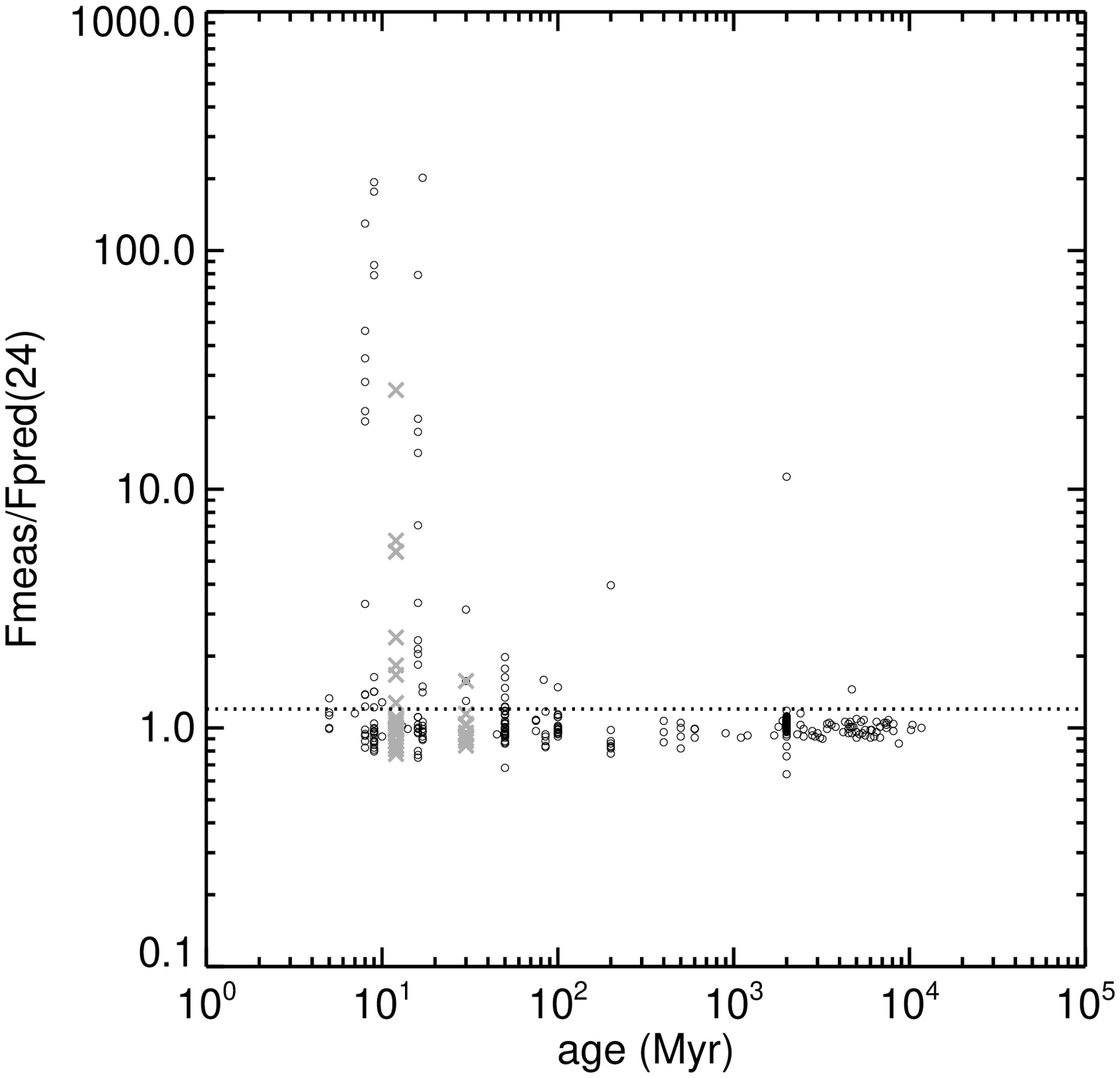}
\caption{The ratio of measured to predicted 24 \mum\ flux as a
function of time for objects in the literature, as described in the
text; gray $\times$ symbols correspond to our objects from
the BPMG and Tuc-Hor. The horizontal dotted line corresponds to the
$F_{\rm meas}/F_{\rm pred}$=1.2 cutoff between disks and photospheres
discussed in \S3.1.}
\label{fig:ff24}
\end{figure*}

\begin{figure*}[tbp]
\epsscale{1.0}
\plotone{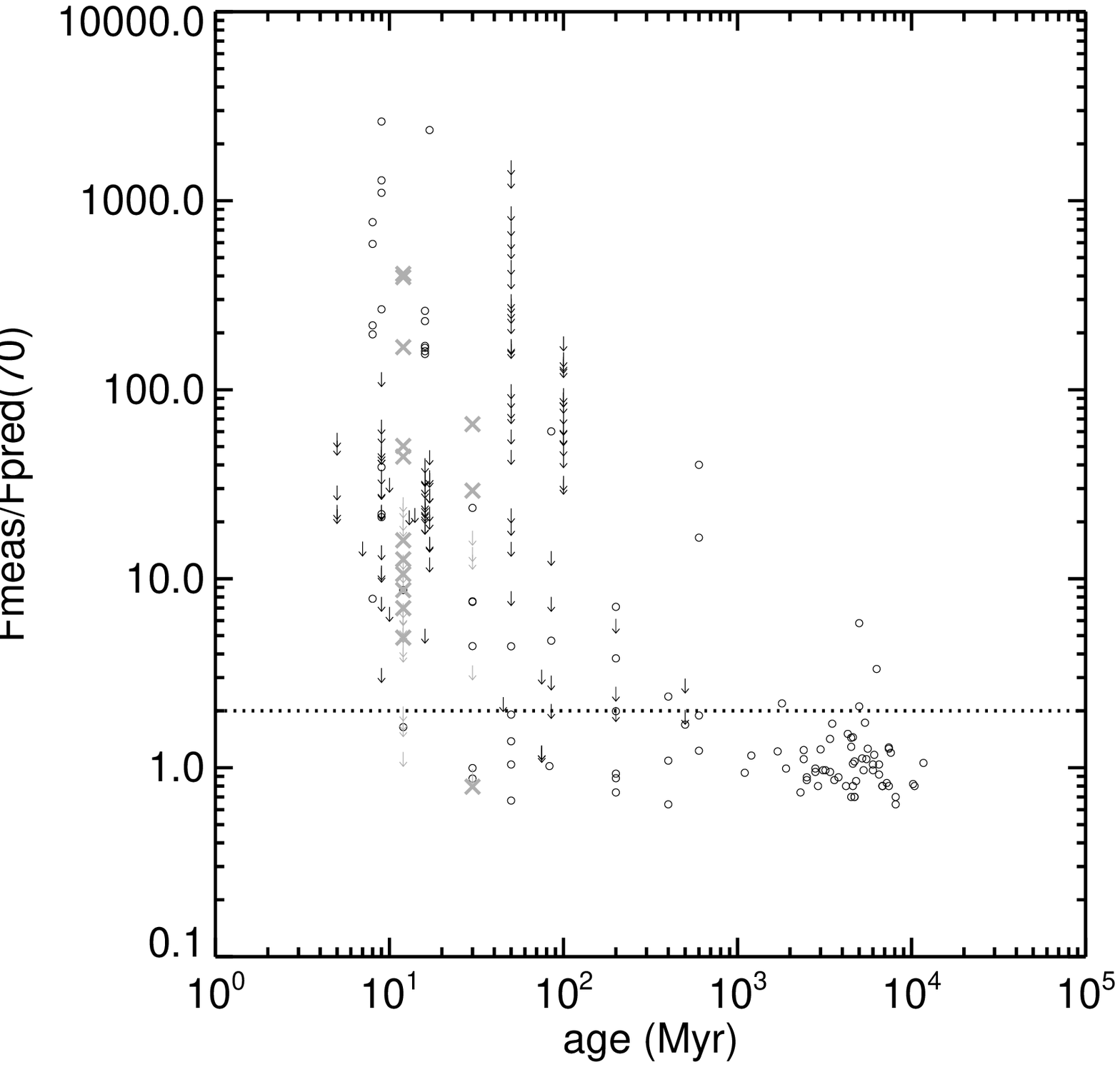}
\caption{The ratio of measured to predicted 70 \mum\ flux as a
function of time for objects in the literature, as described in the
text; gray $\times$ symbols or upper limits correspond to our objects
from the BPMG and Tuc-Hor. The horizontal dotted line corresponds to
the $F_{\rm meas}/F_{\rm pred}$=2 cutoff between disks and
photospheres discussed in \S3.2.   }
\label{fig:ff70}
\end{figure*}
\clearpage

\begin{deluxetable}{lllll}
\tablecaption{Infrared Excess Fractions}
\label{tab:diskfrac}
\tabletypesize{\scriptsize}
\tablewidth{0pt}
\tablehead{
\colhead{cluster/ass'n} & \colhead{age} &\colhead{24 \mum\ disk
fraction}  & \colhead{70 \mum\ disk fraction}
&\colhead{reference}}
\startdata
Upper Sco F\&G & $\sim$5 Myr  & 1/5, 20\% ($\pm$20\%)& 0/5, $>$0\% ($\pm$20\%)& Chen \etal\ (2005a) \\
$\eta$ Cha   & $\sim$8 Myr    & 9/16, 56\% ($\pm$18\%) & 5/15, $>$33\% ($\pm$15\%)& Gautier \etal\ (2008) \\
TW Hya       & $\sim$8 Myr    & 7/23, 30\%\tablenotemark{a} ($\pm$11\%) & 6/20, $>$30\% ($\pm$10\%)& Low \etal\ (2005) \\
UCL \& LCC\tablenotemark{b}~~F\&G & $\sim$10 Myr & 12/35, 34\% ($\pm$10\%)& 7/35, $>$20\% ($\pm$7\%)& Chen \etal\ (2005a) \\
BPMG         & $\sim$12 Myr   & 7/30, 23\% ($\pm$9\%) & 11/30, $>$37\%  ($\pm$11\%)& {\it this work}\\
Tuc-Hor      & $\sim$30 Myr   & 1/9, 11\% ($\pm$11\%) & 8/31, $>$26\% ($\pm$10\%)& Smith \etal\ (2006), \\
              &               &             &               & {\it combined with this work} \\
NGC 2547     & $\sim$25 Myr   & $\sim$25\% & \ldots & Young \etal\ (2004) \\
IC 2391      & $\sim$50 Myr   & 6/26, 23\% ($\pm$9\%) & \ldots & Siegler \etal\ (2006) \\
Pleiades     & $\sim$100 Myr  & 9/54, 17\%  ($\pm$5\%)& none detected & Gorlova \etal\ (2006), \\
              &               &             &               & Stauffer \etal\ (2005)\\
M47          & $\sim$100 Myr  & 8/63, 13\% ($\pm$5\%) & \ldots & Gorlova \etal\ (2004) \\
Hyades       & $\sim$650 Myr  & 0/6, 0\% ($\pm$2\%) & \ldots & Rieke \etal\ (2005) \\
field        & $\sim$4000 Myr & 1/69, 1\% ($\pm$3\%)& 7/69, 10\% ($\pm$4\%)  & Bryden \etal\ (2006) \\
\enddata
\tablenotetext{a}{TWA 24 \mum\ infrared excess fraction reassessed
here; see text for discussion.}
\tablenotetext{b}{UCL=Upper Centaurus-Lupus; LCC=Lower Centaurus-Crux}
\end{deluxetable}
\clearpage

\section{Discussion}
\label{sec:disc}

Based on the standard paradigm, the stars in the BPMG are expected to
have a lower disk frequency and smaller infrared excesses than found
in younger stars, and to possess a higher disk frequency and larger
excesses than older stars.  Our results follow those expectations at
both 24 and 70 \mum; Figure~\ref{fig:evol}  plots our 24 and 70 \mum\
disk (excess) fractions in context with several other determinations
from the literature, which can also be found in Table~5.  After a
brief discussion of some minor issues, we now discuss our study in
context with other studies in the literature.  

Because these disks are likely to evolve such that the infrared
excesses disappear from the ``inside-out'' (\eg, Su \etal\ 2006) it is
important to consider the wavelength dependence of the disk fraction
being considered.  Since the sensitivity of the 70 \mum\ array does
not allow for detections of the stellar photospheres for most stars,
it is difficult to obtain an unambiguous definition of the disk
fraction at this wavelength.  Essentially all studies, therefore,
quote a lower limit to the true 70 \mum\ disk fraction in  clusters or
associations.  The error bars shown in Figure~\ref{fig:evol} and
listed in Table~5 are derived from Poisson (counting) statistics. 
Note too that there are relatively large uncertainties on the ages of
these clusters and associations.  Finally, we note that several of our
stars as considered here are unresolved binaries.  We have made no
attempt to distinguish binaries as a separate population from single
stars here, nor to apportion the flux between the companions, but we
have listed known binarity in Tables 1-3. Given the distance of the
BPMG and the MIPS resolution, unresolved binaries must have a
separation of $\lesssim$ 200 AU. The results of Trilling \etal\ (2007)
suggest that the evolution of such circumbinary disks is roughly
comparable to that of single stars, so including unresolved binaries
as single stars should not significantly change
Figure~\ref{fig:evol}. 

There are three associations in Table~5 thought to be younger than the
BPMG: Upper Sco ($\sim$ 5 Myr), the TW Hydra Association (TWA;
$\sim$8-10 Myr), and the $\eta$ Chamaeleon association ($\sim$5-9
Myr).  All three of these associations have larger 24 \mum\ disk
fractions in the literature (Chen \etal\ 2005a, Low \etal\ 2005,
Gautier \etal\ 2008, respectively) than we find for the BPMG,
consistent with expectations.  (Admittedly, the Upper Sco sample
includes only about 5\% of the likely members of this association, so
there is a large uncertainty on the disk fraction compared to what
future investigators are likely to conclude.)  Low \etal\ (2005) find
for TWA that there are very large excesses around four of the TWA
stars, with possibly a subtle 24 \mum\ excess around one more of the
stars.  We have re-reduced their MIPS data in exactly the same fashion
as here in the BPMG, and find, as did Low \etal, that many of the
measurements are consistent with photospheres.  We were able to
measure 24 \mum\ fluxes for 23 objects, some of which are components
of wide binary systems.  We confirm the 4 large excess objects (TWA 1,
3, 4, 11), as well as the small excess found  in TWA 7, but also,
using the same criteria as for the BPMG, that 8b and 19 are also
likely to harbor circumstellar disks.  Thus, to aid in direct
comparison with our BPMG data, we have taken the TWA disk fraction at
24 \mum\ to be 7/23 stars, or 30\%.  The largest excess objects in TWA
have \ks$-$[24]$>$4 (5.8, 5.0, 4.4, and 4.4 for TWA 1, 3, 4, and 11,
respectively, with F$_{\rm meas}$/F$_{\rm pred}$= 160, 69, 51, and
58).  The reddest object we have is $\beta$ Pic itself, with
\ks$-$[24] of only 3.5, well below the 4 extreme TWA stars.  The three
TWA stars with more moderate excesses, TWA 7, 8b and 19, have
\ks$-$[24]= 0.70, 0.75 and 0.30, respectively.  (The F$_{\rm
meas}$/F$_{\rm pred}$ values we calculate are 1.4, 1.3, and 1.3,
respectively.)  In terms of the 70 \mum\ disk fraction, the numbers
obtained for Upper Sco, TWA, $\eta$ Cha, and BPMG are all consistent,
within 1-$\sigma$ uncertainties, with having a constant disk
fraction.  The one disk candidate from the Chen \etal\ (2005a) Upper
Sco sample has $L_{\rm dust}/L_{*}=4.4\times10^{-4}$.  The values for
$L_{\rm dust}/L_{*}$ for TWA range from 0.27 to $\sim10^{-4}$ (Low
\etal\ 2005), and in $\eta$ Cha, they range from 0.019 to
$\sim10^{-6}$ (Gautier \etal\ 2008); both of these clusters have 
larger $L_{\rm dust}/L_{*}$ values than those we find here in the BPMG
(10-250$\times10^{-5}$).  

The estimated age of the Upper Centarus-Lupus (UCL) and Lower
Centaurus-Crux (LCC) associations has been taken to be $\sim$15-20 Myr
(\eg, Chen \etal\ 2005a), but is more recently set at  $\sim$10 Myr
(Song \etal\ submitted), which we adopt here.  The ages of those
clusters are roughly comparable to that of the BPMG.  Both the 24
and 70 \mum\ disk fractions found in F and G stars from UCL \& LCC are
within 1-$\sigma$ of the disk fractions found in the BPMG, despite the
fact that our BPMG disk fractions include more stars than just F\&G. 
The $L_{\rm dust}/L_{*}$ values found in UCL \& LCC range from
$\sim10^{-3}-10^{-5}$, comparable to the range we find in the BPMG. 

Tucanae-Horologium ($\sim$20-40 Myr) and NGC 2547 ($\sim$25 Myr) are
thought to be slightly older than the BPMG.  Membership in NGC~2547
(Young \etal\ 2004) is not as well-established as it is for other
objects in Table~5.  The 24 \mum\ disk fraction is consistent with
that for the BPMG, and the 70 \mum\ disk fraction is not reported. 
Working in a sample of nearby solar-type young stars (including
several from but not limited to Tuc-Hor), Smith \etal\ (2006) find
that just 19 of their overall 112-star sample (17\%) have 70 \mum\
detections at all.  Of the 22 stars in the Tuc-Hor association
included in the Smith \etal\ sample, 8 are detected, and 6 are
determined to be greater than photospheric, or a lower limit on the
disk fraction of 27\%. We can combine these stars with the 9 Tuc-Hor
stars from the present work, obtaining a 24 \mum\ disk fraction of 1/9
(11\%), and a 70 \mum\ disk fraction of at least 8/31 ($>$26\%). Within
small-number statistics, the Tuc-Hor disk fractions at both 24 and 70
\mum\ are indistinguishable from those obtained here in the BPMG.

There are 4 clusters, in addition to field stars, older than the BPMG
in Table~5. The 24 \mum\ disk fraction reported by Siegler \etal\
(2006) for IC 2391 ($\sim$50 Myr) is comparable to that for the BPMG.
The disk fractions from the Pleiades and M47 ($\sim$100 Myr; Gorlova
\etal\ 2006, 2004) are only marginally lower than that inferred for
the BPMG. The BPMG disk fraction is significantly higher than that for
the Hyades (Rieke \etal\ 2005) or field stars from the solar
neighborhood (Bryden \etal\ 2006).  The Bryden \etal\ (2006) study
found just one 24 \mum\ excess out of 69 stars. Detections (of disks
or photospheres) are harder at the distances of most of these older
clusters; in the Pleiades, no disks are seen at 70 \mum, although the
background is quite high (Stauffer \etal\ 2005).  For the old
($\sim$4000 Myr) field stars in Bryden \etal\ (2006), 10\% of their
$\sim$70 star sample has 70 \mum\ disks.  The $L_{\rm dust}/L_{*}$
values reported by Bryden \etal\ (2006) range from
$<10^{-6}-\sim10^{-5}$, lower than what we find in the BPMG (or even
could have detected). Our results are consistent with the trend that
the disk fraction and brightness falls with time.  

In considering these disk fractions, we have grouped together stars of
a range of masses in order to increase the number of stars considered
at each age; for example, the BPMG disk fraction includes stars from A
to M.  However, disk evolution is probably stellar-mass-dependent
(e.g., Carpenter \etal\ 2006), and certainly measured colors are
mass-dependent (as discussed above; see Figure~\ref{fig:kk24_spty}).  
Besides $L_{\rm dust}/L_{*}$, another way that we might attempt to
compensate for the range of spectral types is to use the ratio of
measured to predicted flux densities.    Figure~\ref{fig:ff24} and
\ref{fig:ff70} present the ratios of predicted to measured flux
densities for 24 and 70 \mum\ for our stars and, where possible,
values from the literature for individual stars (Bryden \etal\ 2006; 
Chen \etal\ 2005a,b; Gautier \etal\ 2008; Kim \etal\ 2005; Low
\etal\ 2005; Siegler \etal\ 2007; Smith \etal\ 2006; Stauffer \etal\
2005).   Where previous work has not reported a predicted flux density
for each star, we have calculated the predicted flux densities by the
same methodology as above for each star (finding the nearest grid
point in the Kurucz-Lejeune model grid for a given spectral type and
interpolating to the MIPS effective wavelengths).  The upper envelope
found in these figures is similar to the 24 \mum\ upper envelope found
by Rieke \etal\ (2005) or Su \etal\ (2006) for 24 \mum\ excesses
around A stars, or at 70 \mum\ by Su \etal\ (2006).  The range of
excess strengths found at any age could be a result of initial
conditions, rates of evolution, or recent collisional events; there is
no obvious way to determine the origin from these data alone. Currie
\etal\ (2008) report seeing the decline of primordial disks and the
rise of debris disks; this reinforces the importance of further study
of stars with a range of excesses in the 8-10 Myr age range,
specifically the need for high-quality complete disk fractions. 

\clearpage
\begin{figure*}[tbp]
\epsscale{1.0}
\plotone{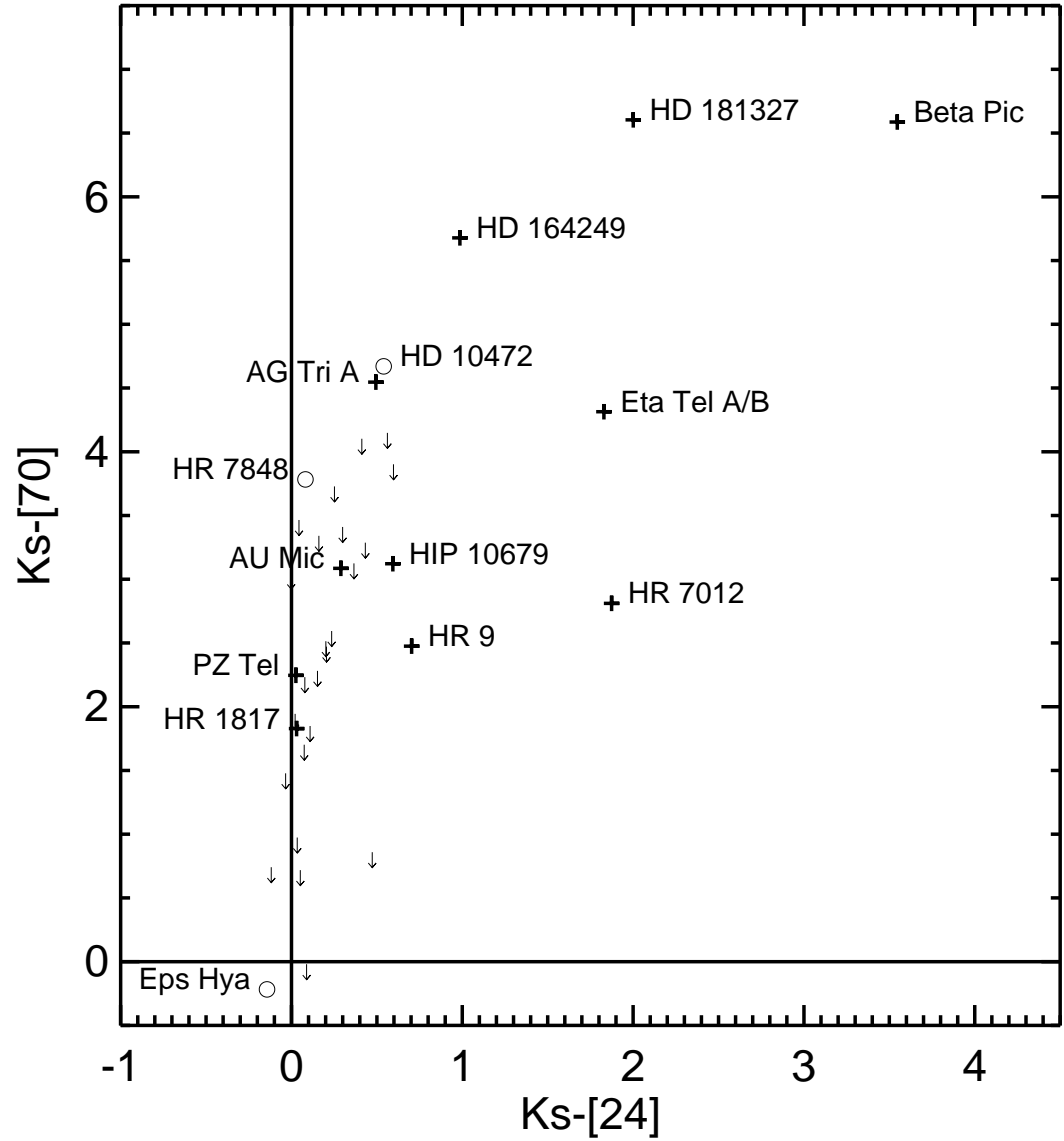}
\caption{Plot of \ks$-$[24] vs.\ \ks$-$[70] for all of the objects
considered here.  Plus signs are objects detected (at 70 \mum) from
the BPMG; open circles are detected objects from Tucanae-Horologium.
All other objects (from both associations) are indicated as upper
limits at 70 \mum.  These results are consistent with an
``inside-out'' infrared excess reduction scenario, where 24 \mum\
excesses disappear before 70 \mum\ excesses; see text for further
disucssion.}
\label{fig:colorcolor}
\end{figure*}
\clearpage

Figure~\ref{fig:colorcolor} shows \ks$-$[24] vs.\ \ks$-$[70] for the
objects considered here. It is clear not only which stars with
excesses in one band also have excesses in the other band, but also
very roughly the correlation of the size of the excess (with all the
caveats about spectral type dependence discussed above).  The MIPS
measurements of $\epsilon$ Hya are consistent with a purely
photospheric origin for its IR flux.  Of the 8 stars identified above
as having any excesses at 24 \mum, all also have clear excesses at 70
\mum.  All four objects with the largest 24 \mum\ excesses also have
large 70 \mum\ excesses.   Five additional stars are detected as
having excesses at 70 \mum, but without significant excesses at 24
\mum.  For the stars with disk excesses at 24 \mum, the median
\ks$-$[24] is 0.99 magnitudes; for those same stars, the median
\ks$-$[70] is 4.5 magnitudes, significantly redder.  

A disk may be inferred to have an inner hole if it has an infrared
excess at long wavelengths but not at short wavelengths, such as these
stars with signficant 70 \mum\ excess and very small 24 \mum\ excess. 
By this definition, the majority of debris disks around older main
sequence FGK stars possess inner holes (29 of 37 disks;  Trilling
\etal\ 2008), whereas only 8/44 debris disks around younger A stars do
(Su \etal\  2006).  At ages of a few Myr, the circumstellar disks
found in star-forming regions have a very low MIPS inner hole
frequency (Rebull \etal\ 2007; Harvey \etal\ 2007; Young \etal\
2005).  MIPS studies of young associations such at the BPMG provide a 
key bridge between the massive, young disks that generally lack inner
holes, and  the older, tenuous debris disks that often possess them. 
At age 8 Myr, the TW Hya and $\eta$ Cha groups show very few disks
with MIPS inner holes (1/6 disks from TWA, Low \etal\ 2005 and
reduction above, and 0/5 disks from $\eta$ Cha, Gautier \etal\ 2008).
These young associations also possess a mixed population of disks
with  fractional infrared luminosities near 0.1 (characteristic of
massive primordial disks, such as that of TW Hya) and $<$0.001
(characteristic of optically thin debris disks, such as that of
$\beta$ Pictoris).  None of the stars with disks in the larger Sco-Cen
association (part of which is age $\sim$5 Myr and the rest of which is
age $\sim$10 Myr) possess MIPS inner holes (Chen \etal\ 2005a).  The
12 Myr old BPMG (this work) contains only optically thin disks, with
4/11 disks possessing MIPS inner holes (note that we are including AG
Tri A, since it has a proportionally much larger 70 \mum\ excess than
any potential small 24 \mum\ excess).  In the  $\sim$ 30 Myr old
Tuc-Hor association, 6/8 stars with disks have inner holes (this work,
combined with Smith \etal\ 2006). A smooth increase of inner hole
frequency with time is evident, and although small number statistics
prevent strong conclusions, it is clear that the BPMG is the youngest
stellar group in which the frequency of MIPS inner holes is clearly
larger than that seen in the pre-main sequence stellar population.
What is seen in the BPMG and these other clusters is consistent with
expectations based on other clusters that stars lose their 24 \mum\
excesses before their 70 \mum\ excesses (``inside-out''; \eg, Su
\etal\ 2006).    

\clearpage
\begin{deluxetable}{ll}
\tablecaption{\vsini\ Values Used for BPMG Stars G0 and Later}
\label{tab:vsini}
\tablewidth{0pt}
\tablehead{
\colhead{star} & \colhead{\vsini\ (km s$^{-1}$)} }
\startdata
HIP 10679       &   7.8                 \\
HIP 12545       &   9.3                 \\
GJ 3305         &   5.3                 \\
HIP 23309       &   5.8                 \\
GJ 3322 A/B     &   7.7                 \\
AO Men          &  16                   \\
V343 Nor A/B    &  11                   \\
V824 Ara A/B    &  37 (companion 34)    \\
CD-64D1208 A/B  & 102.7                 \\
PZ Tel          &  63                   \\
AT Mic A/B      &  10.6 (companion 17) \\
AU Mic          &   8.5                 \\
AZ Cap A/B      &  14.6                 \\
WW PsA A        &  14.0                 \\
WW PsA B        &  24.3                 \\
\enddata
\end{deluxetable}
\clearpage

\begin{figure*}[tbp]
\epsscale{1.0}
\plotone{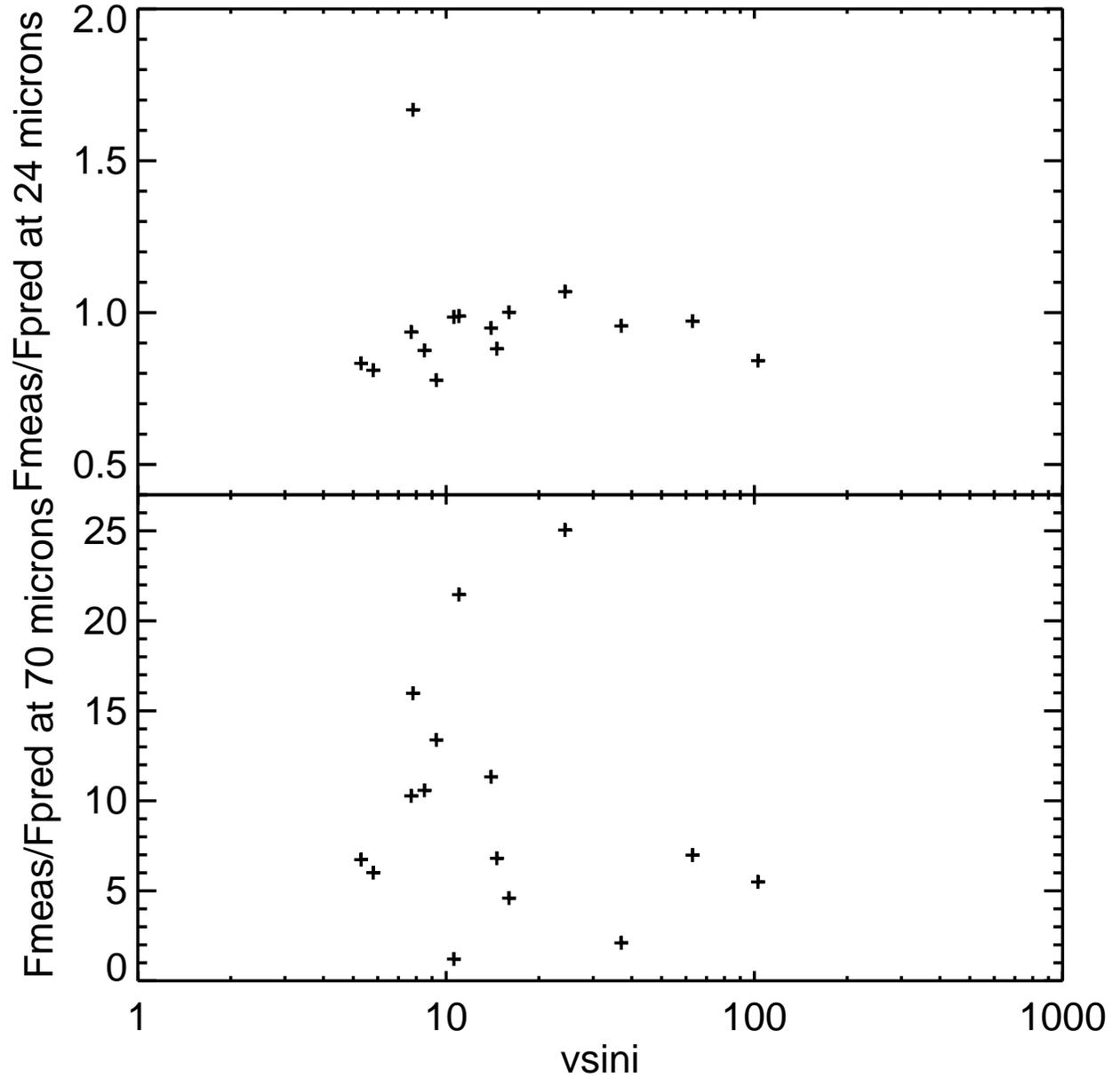}
\caption{Plot of \vsini\ (in km s$^{-1}$) vs.\ F$_{\rm meas}$/F$_{\rm
pred}$ at 24 \mum\ (top) and at 70 \mum\ (bottom) for all of the BPMG
G, K, and M stars considered here.  While certainly not conclusive,
these figures are reminiscent of effects seen in younger clusters such
as Orion.}
\label{fig:rotation}
\end{figure*}
\clearpage

The G, K, and M stars in at least some clusters that are much younger
than the BPMG, $\sim$1-5 Myr old, exhibit a correlation between
rotation and infrared excess in that slower rotators are more likely
to have infrared excesses, or disks (see, \eg, Rebull \etal\ 2006 and
references therein).  This agrees with theoretical expectations in
that the young lower-mass GKM stars are thought to have strong
magnetic fields that thread the (primordial) circumstellar disk,
mediating accretion and locking the rotation of the star to that of
the disk. However, by the $\sim$12 Myr age of the BPMG, and at the
distances from the parent star of these disks emitting at 24 and 70
\mum, no disk locking is expected to still be operating.  In
Figure~\ref{fig:rotation}, we examine the correlation of disk excess
with rotation rates for the G, K, and M BPMG members. (The \vsini\
values used for these stars appear in Table~6.) The faster-rotating
lower-mass stars in the BPMG in Figure~\ref{fig:rotation} show a {\em
weak} tendency to have a smaller disk excess.   While certainly not
conclusive, these figures are suggestive.  Additional \vsini\ and
rotation period determinations would be useful to test this
correlation, as well as additional Spitzer measurements in other
similarly aged clusters.   Interestingly, Stauffer \etal\ (2007) find
a similar correlation between 24 \mum\ excess and \vsini\ seen in open
clusters primarily from the FEPS program (Formation and Evolution of
Planetary Systems; Meyer \etal\ 2006) and the Pleiades.  Given that
all of our disk candidates in the BPMG now posess at best tenuous
debris disks, the disk mass is insufficient (now) to regulate the
stellar angular momentum as in the case of massive primordial disks. 
Perhaps these disks started out as more massive than the other BPMG
members.  Perhaps the disk dispersion timescale, which determines
whether or not a disk still persists at $\sim$12 Myr, is set early on
in the lifetime of the disk, when the angular momentum and mass flux
through the disk is the highest, the central object is large, and the
influence of disk locking (or braking) is the strongest.  In that
case, the {\em weak} correlation seen in Figure~\ref{fig:rotation} is the
signature of a process operating at earlier times. 

Alternatively, strong stellar winds could play an important role in
clearing the disk of small particles, as suggested by Plavchan \etal\
(2005) and Chen \etal\ (2005a).  Rapid rotation, which enhances the
stellar dynamo and presumably the strength of the stellar wind, would
then be associated with more tenuous disks as suggested by the data in
Figure~\ref{fig:rotation}.  Wind ablation of dust could be an ongoing
process. 

We see no obvious way to test for whether winds (operating now and/or
in the past) or disk locking (operating in the past) are more likely
using these data; clearly these initial results will need future
observational follow-up, such as the use of periods rather than
\vsini, and a search for similar effects in other similarly-aged
clusters.  We emphasize again for clarity that the correlation seen in
Figure~\ref{fig:rotation} is only for the GKM stars in the BPMG.

\section{Conclusions}
\label{sec:concl}

We have presented here MIPS 24 and 70 \mum\ observations of 30 stars
or star systems in the BPMG, as well as nine from Tucanae-Horologium,
with 160 \mum\ observations for a subset of 12 BPMG stars.   In
several cases, the new MIPS measurements resolve source confusion and
background contamination issues in the previous IRAS data. 

We found that 7 BPMG members have signficant 24 \mum\ excesses, or a
disk fraction of 23\%.  Eleven BPMG systems have significant 70
\mum\ excesses (disk fraction of  $\geq$37\%, as this is a lower
limit).  Five exhibit 160 \mum\ excesses, out of a biased sample of 12
observed, and they have a range of 70:160 micron flux ratios.  The
disk fraction, and the size of the excesses measured at each
wavelength, are both consistent with an ``inside-out'' infrared excess
reduction scenario, wherein the shorter-wavelength excesses disappear
before longer-wavelength excesses, and consistent with the overall
decrease of disk frequency with stellar age, as seen in Spitzer
studies of other young stellar groups.  

We characterized the disk properties using simple models and
fractional infrared luminosities.  Optically thick disks, seen in the
8 Myr age TW Hya and $\eta$ Cha associations, are entirely absent in
the BPMG at age 12 Myrs.

\acknowledgements 

L.\ M.\ R.\ wishes to acknowledge funding from the Spitzer
Science Center to allow her to take a ``science retreat'' to
work intensively on this paper.   
The authors wish to acknowledge the MIPS GTO team for allowing us to
use the DAT to process the 160  \mum\ data. This work is based on
observations made with the Spitzer Space Telescope, which is operated
by the Jet Propulsion Laboratory, California Institute of Technology
under a contract with NASA.  Support for this work was provided by
NASA through an award issued by JPL/Caltech.   This research makes use
of data archived and served by the NASA Star and Exoplanet Database
(NStED) at the Infrared Processing and Analysis Center. NStED is
jointly funded by the National Aeronautics and Space Administration
(NASA) via Research Opportunities in Space Sciences grant 2003 TPF-FS,
and by NASA's Michelson Science Center. NStED is developed in
collaboration with the NASA/IPAC InfraRed Science Archive (IRSA). This
research has also made use of NASA's Astrophysics Data System (ADS)
Abstract Service, and of the SIMBAD database, operated at CDS,
Strasbourg, France.  This research has also made use of data products
from the Two Micron All-Sky Survey (2MASS), which is a joint project
of the University of Massachusetts and the Infrared Processing and
Analysis Center, funded by the National Aeronautics and Space
Administration and the National Science Foundation.  These data were
served by the NASA/IPAC Infrared Science Archive, which is operated by
the Jet Propulsion Laboratory, California Institute of Technology,
under contract with the National Aeronautics and Space
Administration.  The research described in this paper was partially
carried out at the Jet Propulsion Laboratory, California Institute of
Technology, under contract with the National Aeronautics and Space
Administration.  

\appendix
\section{Comments on individual objects}
\label{sec:indobj}

These comments on individual objects address the issues of (possibly)
resolved objects, serendipitous detections, IR cirrus, and multiple
systems.  In some cases, the proximity of a true companion and/or
infrared cirrus results in the low-spatial-resolution IRAS fluxes
being anomalously high when compared with the MIPS fluxes. All of
those instances are discussed here.  

In several cases, objects in close proximity to the target object were
detected.  Since these objects are bright enough to be detected in
these shallow observations, these additional objects are also
potential association members, and/or contributors to source confusion in
lower spatial resolution observations such as IRAS.  Based on the MIPS
measurements, we conclude none are association members; see individual
discussion below.

\subsection{HIP 3556 (Tuc-Hor)}

At 24 \mum, there are several objects easily visible besides the
target, with several being of comparable brightness to the target. 
Two of them are easily visible in the 70 \mum\ image whereas HIP 3556
is undetected. Few of them have obvious counterparts in a POSS or
2MASS image.  Given their evidently steeply rising SEDs, we suspect
that they are background galaxies.

\subsection{$\phi$ Eri (Tuc-Hor)}

Spitzer observations of $\phi$ Eri clearly detect it in 24 \mum\ to be
173 mJy; there is an emission peak at this location at 70 \mum, but it
is comparable in size to the noise fluctuations found in this region,
so it is listed as an upper limit in our study.  The upper limit falls
right on the expected photospheric flux.

There is a nearby source 90$\arcsec$ away at 02h16m30.6s, -51d30m44s,
measured to be 12.3 mJy (at 24 \mum) .  This object is not detected at
70 \mum, but it is detected in 2MASS with \ks=4.13 mag.  The resultant
\ks$-$[24] color suggests that it is far too blue to be a star, but
the PSF as seen in POSS plates appears stellar.  This source is
probably not a new association member.

\subsection{HD 14082 and HIP 10679}

HIP 10679 and HD 14082 are close enough to each other
($\sim10\arcsec$) to be observed in the same MIPS photometry field of
view.  Both objects are point sources at 24 \mum\ and have comparable
fluxes at this bandpass.  At this separation, these objects should be
distinguishable at 70 \mum, but only one object is detected.  Based on
the central position of the object, we have assigned the measured flux
to HIP 10679.  This is a weak detection, with a signal-to-noise ratio
of only $\sim$5.  The PSF appears to be elliptical, with the major
axis roughly a factor of twice the minor axis.  It is not extended in
the direction of the companion, or in the direction of the scan mirror
motion.  While it is possible that the object is truly resolved at 70
\mum, the fact that it is not resolved at 24 \mum\ leads us to suspect
that the apparently elliptical PSF is instrumental in nature.  The
object is so faint as to not be easily detectable in subsets of the
data, so it is difficult to assess whether or not co-adding the data
has caused this effect.

\subsection{GSC 8056-482 (Tuc-Hor)}

While only one BPMG object is expected to be included in this
observation, several fainter objects are clearly detected in the 24
\mum\ image.  The closest and brightest one to GSC 8056-482 is
23$\arcsec$ away, located at  02h36m49.1s, -52d03m12.3s, and is
measured to be 2.0 mJy.  It is not detected in 2MASS or at 70 \mum.

\subsection{HR 6070}

HR 6070 appears in the IRAS PSC (but not the FSC) as a detection at
all IRAS bands with coordinates slightly offset to the northwest from
the optical position.  However, the MIPS observations reveal an
isolated point source with lower flux measured at 24 \mum\ and an
upper limit at 70 \mum\ that is comparable to the detection reported
by IRAS.  The 24 \mum\ image reveals clear cirrus on the northwest
side of the image, in the same direction as the reported center of the
IRAS source, suggesting that the measured IRAS flux is contaminated by
infrared cirrus.  If all of the flux attributed to the point source
in the IRAS catalog were really coming from the point source, we would
have detected it, but we did not.  The MIPS observations provide a
much better understanding of any infrared excess present in this star,
suggesting no excess at 24 \mum\ and providing a constraint at 70
\mum.

\subsection{V824 Ara A, B, \& C}
\label{sec:v824}

This triple system, located all within an arcminute, was also
unresolved by IRAS.  MIPS can clearly separate C from A/B at 24 \mum,
but no objects are detected at 70 \mum.  IRAS's beam size encompasses
all three of these components. MIPS resolves the source confusion and
does not find an IR excess in A/B or C.  

In addition to the components of this system, MIPS sees two additional
objects, neither of which are seen at 70 \mum.  Neither of these
objects have a \ks$-$[24] color suggestive of an excess.

\subsection{HD 164249}

In the 24 \mum\ image for HD 164249, three objects are present, two of
which are also seen at 70 \mum.  (None of the objects are seen in our
160 \mum\ data.) The target of the observation is clearly apparent in
both 24 and 70 \mum, and a second object appears 0.76$\arcmin$ away
with a 24 \mum\ flux of 1220 mJy and a 70 \mum\ flux of 172 mJy.  A
third faint object 1.4$\arcmin$ away has a 756 $\mu$Jy flux at 24
\mum. Both of these objects appear in the 2MASS catalog.  The brighter
object has a \ks$-$[24] color of 7.4; the fainter object has
\ks$-$[24]=0.002.  The latter is a photosphere with arguably no excess
whatsoever at 24 \mum. This, combined with its overall faintness,
suggests it is probably a background star.  The former appears as a
very faint smudge on POSS plates and has a clear elliptical shape in
2MASS images.  The object appears in the 2MASS extended source catalog
as a galaxy with name 2MASXJ18030752-5139225.  It likely has
influenced the measured flux for HD 164249 in lower spatial resolution
measurements.

\subsection{HR 6749/HR 6750}

This binary system is unresolved by MIPS.  IRAS measures a detection
at all 4 bands suggesting an infrared excess and therefore
circumstellar dust.  MIPS is able to resolve apparent source
confusion, placing the 24 \mum\ point at a photospheric level and
putting constraints on the 70 \mum\ flux.  The 24 \mum\ image suggests
that there may be infrared cirrus that contributed to the measured
IRAS flux; any background flux is not very bright at MIPS-24 (while
MIPS has much more sensitive detectors that IRAS, it also samples much
smaller angles on the sky, so the surface brightness sensitivity is
not substantially different than IRAS).  At 70 \mum, if all
of the flux attributed to the point source in the IRAS catalog were
really coming from the point source, we would have detected it, but we
did not.

\subsection{AT Mic}

This object is detected at 24 \mum, but not at 70 \mum. 
There is another object at 24 \mum\ that is 1.5$\arcmin$
away at 20h41m55.4s, $-$32d24m57s, with a 24 \mum\ flux of 2.8
mJy.  This object has a \ks$-$[24] color of $-$0.02, which is
not indicative of any excess.

\subsection{Resolved objects}

We note for completeness that at least three objects in the BPMG,
$\beta$ Pic itself (\eg, Golimowski \etal\ 2006), AU Mic (\eg, Graham
\etal\ 2007), and HD 181327 (\eg, Schneider \etal\ 2006), are known to
be resolved at other wavelengths.  AG Tri A may be resolved as well
(Ardila \etal\ 2007 in prep).  Of these, $\beta$ Pic itself is the
only one known to be resolved at MIPS wavelengths (Su \etal\ in
preparation, see also Chen \etal\ 2007); the others, if they are
resolved at MIPS wavelengths, are only subtly larger than the
instrumental PSF.  All of these famous objects are extensively
discussed elsewhere, so we do not discuss them again here.

\end{document}